\documentclass[prd,notitlepage,longbibliography,nofootinbib,superscriptaddress,onecolumn,preprintnumbers]{revtex4-2}
\usepackage[utf8]{inputenc}

\usepackage{bm}
\usepackage{comment} 
\usepackage[colorlinks=true,urlcolor=blue,anchorcolor=black,citecolor=blue,linkcolor=red,filecolor=black,menucolor=black,pagecolor=black,linktocpage=true,pdfproducer=medialab,pdfa=true]{hyperref}
\usepackage{graphicx}
\usepackage{amsmath,latexsym,amssymb,mathrsfs,ascmac,mathtools}
\usepackage{multirow}
\usepackage{braket}
\usepackage{placeins}
\usepackage{float}
\usepackage{subcaption}
\usepackage{makecell}
\begin{document}

\title{Distinguishing Dymnikova and Schwarzschild Black Holes through Gravitational Lensing with EFT-Corrected Photon Propagation}

\author{Takamasa Kanai}
\email{kanai@kochi-ct.ac.jp}

\affiliation{Department of Social Design Engineering,
National Institute of Technology (KOSEN), Kochi College,
200-1 Monobe Otsu, Nankoku, Kochi, 783-8508, Japan}

\begin{abstract}
We investigate whether a Dymnikova regular black hole can be observationally distinguished from a Schwarzschild black hole through strong gravitational lensing, taking into account effective field theory (EFT) corrections to photon propagation. We derive the modified photon propagation law induced by non-minimal couplings between the electromagnetic field and spacetime curvature and analyze the resulting photon trajectories in the Dymnikova spacetime. Within the strong deflection limit, we derive the EFT corrections to the photon sphere and the strong-deflection coefficients characterizing the logarithmically divergent behavior of the deflection angle. Although the EFT corrections are parametrically small, their effects can become relevant in the strong-deflection regime, where the deflection angle is highly sensitive to photon propagation near the critical orbit. We evaluate the strong-deflection observables for representative values of the Dymnikova parameter $\ell$ and compare them with the corresponding Schwarzschild results. We find that EFT corrections to photon propagation leave characteristic imprints on the strong-lensing observables. Furthermore, the contribution of the EFT corrections becomes more pronounced as the Dymnikova parameter $\ell$ approaches its critical value, indicating that the near-critical regime provides a particularly sensitive setting in which curvature-dependent corrections to photon propagation may affect gravitational lensing. These results suggest that strong gravitational lensing, together with EFT-induced modifications of photon propagation, may provide a means of distinguishing Dymnikova regular black holes from Schwarzschild black holes.
\end{abstract}
\maketitle

\section{Introduction}

Gravitational phenomena in the strong-field regime provide a unique window into the nature of gravity and the structure of compact objects beyond the weak-field approximation. In particular, the propagation of light near unstable photon orbits gives rise to characteristic observables such as black hole shadows \cite{Falcke:1999pj,EventHorizonTelescope:2019dse,EventHorizonTelescope:2022wkp,Vagnozzi:2022moj} and gravitational lensing \cite{Virbhadra:1999nm,Bozza:2001xd,Bozza:2002zj,Gibbons:2008rj}. These observables provide promising probes for testing the nature of compact objects and searching for deviations from the Schwarzschild black hole predicted by general relativity.

Among these probes, strong gravitational lensing is particularly sensitive to the geometry in the vicinity of an unstable photon orbit. In the strong deflection limit, the deflection angle develops a logarithmic divergence as the closest approach radius approaches the critical radius associated with the unstable photon orbit. This behavior is captured by the formalism developed by Bozza \cite{Bozza:2002zj,Bozza:2002af,Bozza:2003cp}, in which the deflection angle is characterized by a logarithmically divergent term and a finite regular contribution. Since the coefficients of this expansion are determined by the local properties of the effective geometry near the critical orbit, strong gravitational lensing can be particularly sensitive to small deviations from the underlying spacetime geometry.

This sensitivity makes strong gravitational lensing a useful probe of alternative compact-object geometries, including regular black holes \cite{1968qtr..conf...87B,Ayon-Beato:1998hmi,Ayon-Beato:1999kuh,Hayward:2005gi,Ashtekar:2005qt,Modesto:2005zm}. Regular black holes provide phenomenological models in which the central curvature singularity is avoided while a black-hole-like exterior is retained. Their strong-field geometries can therefore differ from the Schwarzschild geometry, leading to modifications of the properties of unstable photon orbits and the associated gravitational-lensing observables. Among the various regular black-hole solutions, the Dymnikova black hole provides a representative example of a nonsingular black-hole spacetime with a Schwarzschild-like asymptotic structure. The parameter characterizing the Dymnikova geometry modifies the strong-field region and consequently affects the photon sphere, critical impact parameter, and strong-deflection behavior. This naturally raises the question of whether a Dymnikova regular black hole can be observationally distinguished from the Schwarzschild black hole through strong gravitational lensing.

The propagation of photons, however, need not be determined solely by the background spacetime geometry. Within the effective field theory (EFT) framework \cite{Weinberg:1978kz,Donoghue:1993eb,Donoghue:1994dn,Burgess:2003jk}, higher-dimensional operators can generate non-minimal couplings between the electromagnetic field and spacetime curvature. Such curvature-photon interactions modify the photon dispersion relation and, consequently, the propagation of light in curved spacetime. In particular, these corrections can depend on the photon polarization and give rise to gravitational birefringence \cite{Scharnhorst:1990sr,Barton:1989dq,Barton:1992pq,Latorre:1994cv,Dittrich:1998fy,DeLorenci:2000yh,Drummond:1979pp,Daniels:1993yi,Shore:1995fz,Daniels:1995yw,Shore:2007um,Cho:1997vg,Izumi:2014loa,Reall:2014pwa,Allahyari:2019jqz,Cao:2021sty,Reall:2021voz,Davies:2021frz,Fu:2025oxr,Kanai:2026ohg}. As a result, the photon trajectories relevant for gravitational lensing can deviate from the null geodesics of the background metric. The strong-deflection observables are therefore sensitive not only to the geometry of the compact object but also to the curvature-dependent dynamics governing photon propagation. This provides an additional channel through which different compact-object spacetimes may be distinguished.

The Dymnikova spacetime is particularly well motivated in this context because, unlike conventional four-dimensional polynomial quasitopological gravity \cite{Bueno:2024dgm,Oliva:2010eb,Myers:2010ru,Dehghani:2011vu,Cisterna:2017umf,Bueno:2019ltp,Bueno:2019ycr,Bueno:2022res,Bueno:2024zsx,Konoplya:2024kih,Bueno:2025qjk,Bueno:2025tli,Ling:2025ncw,Frolov:2025ddw,Frolov:2026rcm}, nonpolynomial quasitopological gravity has been investigated as a framework for constructing regular black-hole solutions~\cite{Deser:2007za,Gao:2012fd,Colleaux:2015yta,Colleaux:2017ibe,Colleaux:2019ckh,Bueno:2025zaj,Tan:2025hht,Borissova:2026wmn}, and the Dymnikova spacetime is known to be realizable as a vacuum solution of a four-dimensional pure-gravity theory within this framework~\cite{Borissova:2026wmn}. Quasi-topological gravity provides a class of higher-curvature theories in which the curvature invariants are constructed to yield simplified gravitational field equations for highly symmetric spacetimes \cite{Bueno:2025zaj,Borissova:2026wmn}. In four dimensions, the conventional polynomial quasi-topological constructions do not lead to the required nontrivial vacuum dynamics, whereas their nonpolynomial extension admits the Dymnikova geometry as a vacuum solution. Thus, the Dymnikova spacetime can be viewed not merely as a phenomenological regular black-hole metric, but as a geometry arising from a well-defined higher-curvature gravitational framework.

A particularly relevant feature of the Dymnikova spacetime is that, unlike the Schwarzschild vacuum, it is generally not Ricci-flat. Consequently, curvature-photon interactions involving the Ricci tensor can contribute to photon propagation in the Dymnikova background, whereas they vanish in the Schwarzschild spacetime. At the same time, the Weyl curvature contributes through the corresponding curvature-photon interaction in both spacetimes. The EFT corrections therefore provide an additional probe of the curvature structure of the Dymnikova spacetime beyond the modifications encoded in the background metric. This distinction is especially relevant when comparing the Dymnikova and Schwarzschild cases, since the two spacetimes differ not only in their strong-field metric structure but also in the curvature directly experienced by propagating photons. We therefore investigate strong gravitational lensing in the Dymnikova spacetime while consistently taking into account both the background geometry and the EFT corrections to photon propagation.

Although the EFT corrections to photon propagation are perturbatively small, their effects on strong-lensing observables need not be negligible. In the strong-deflection regime, the logarithmic divergence of the deflection angle reflects the prolonged propagation of photons in the vicinity of the critical photon orbit, enhancing the sensitivity of lensing observables to small changes in the photon-sphere structure and photon propagation law. This sensitivity can be particularly pronounced in the near-critical regime of the Dymnikova spacetime, where the properties of the photon orbit change significantly as the Dymnikova parameter approaches its critical value. Strong gravitational lensing therefore provides a useful setting in which to investigate the interplay between the background geometry and curvature-dependent corrections to photon propagation, and may offer a means of distinguishing Dymnikova regular black holes from the Schwarzschild black hole.

In this work, we investigate strong gravitational lensing by a Dymnikova regular black hole, taking into account both the background geometry and EFT corrections to photon propagation. We first analyze photon propagation and strong gravitational lensing in the Dymnikova spacetime in the absence of EFT corrections. We derive the photon-sphere radius, critical impact parameter, deflection angle, and the associated strong-deflection coefficients, and establish the dependence of these quantities on the Dymnikova parameter. We then incorporate the curvature-dependent EFT corrections and derive the modified photon propagation equations together with the corresponding effective optical metrics for the physical polarization modes. Since the EFT corrections depend on the photon polarization, the resulting photon trajectories, critical photon orbits, and strong-deflection coefficients acquire polarization dependence.

We evaluate the strong-deflection observables for representative values of the Dymnikova parameter and compare them with the corresponding Schwarzschild results. We focus in particular on the photon-sphere properties, deflection angle, and photon propagation time, and examine how they are modified by the EFT corrections and how these modifications depend on the Dymnikova parameter. This allows us to assess separately the effects arising from the difference in the background geometry and those induced by the curvature-dependent modification of photon propagation. Our analysis demonstrates that strong gravitational lensing can provide a sensitive probe of both the non-Schwarzschild geometry of the Dymnikova spacetime and the EFT corrections to photon propagation, particularly in the near-critical regime of the Dymnikova parameter.

In Sec.~\ref{sec:DB}, we review the Dymnikova black hole spacetime, introducing its metric and discussing its basic properties, including the photon-sphere structure and the critical parameter of the Dymnikova geometry. In Sec.~\ref{sec:photon}, we introduce the curvature-photon couplings in the effective field theory framework and derive the modified photon propagation law in the Dymnikova spacetime. We then construct the corresponding effective optical metrics for the physical polarization modes. In Sec.~\ref{sec:strong lens}, we investigate gravitational lensing in the strong-deflection limit by taking into account the EFT corrections to photon propagation. We derive the corrections to the photon-sphere properties and the strong-deflection expansion of the deflection angle, and examine their dependence on the Dymnikova parameter and the photon polarization. In Sec.~\ref{sec:time delay}, we study the photon time delay in the strong-deflection regime, consistently incorporating the EFT corrections to photon propagation, and investigate its dependence on the Dymnikova parameter and the polarization modes. Finally, in Sec.~\ref{sec:conclusion}, we discuss the physical implications of our results, compare the Dymnikova and Schwarzschild black holes, summarize our main findings, and discuss possible directions for future work.

In this paper, we set the Newton constant and the speed of light equal to unity.

\section{Dymnikova Spacetime from Higher-Curvature Gravity}
\label{sec:DB}

Regular black-hole spacetimes can arise as vacuum solutions of gravitational theories containing higher-curvature corrections to General Relativity. In particular, Konoplya and Zhidenko demonstrated that the $D$-dimensional extension of the Dymnikova black hole can be obtained within a framework involving an infinite tower of quasi-topological higher-curvature terms~\cite{Konoplya:2024kih}. We consider the action
\begin{align}
S_{\rm QT}=\frac{1}{16\pi G}\int d^D x \sqrt{|g|}\mathcal{L}(R,R_{\mu\nu},C_{\mu\nu\rho\sigma}),
\label{action 1}
\end{align}
where $\mathcal{L}$ is a nonpolynomial function of the curvature tensors, and $\mathcal{L}(R,R_{\mu\nu},C_{\mu\nu\rho\sigma})$ represents the corresponding nonpolynomial quasi-topological curvature Lagrangian. For a static and spherically symmetric spacetime,
\begin{align}
ds^2=-f(r)dt^2+\frac{dr^2}{f(r)}+r^2d\Omega_{D-2}^2,
\end{align}
it is convenient to introduce
\begin{equation}
f(r)=1-r^2\psi(r).
\end{equation}
The corresponding field equations can then be reduced to
\begin{equation}
\frac{d}{dr}
\left[
r^{D-1}h(\psi(r))
\right]
=0,
\end{equation}
where the characteristic function $h(\psi)$ is given, for a polynomial higher-curvature theory, by
\begin{equation}
h(\psi)=\psi+\sum_{n=2}^{n_{\max}}\alpha_n\psi^n.
\end{equation}
Consequently,
\begin{equation}
h(\psi(r))=\frac{\mu}{r^{D-1}},
\end{equation}
with $\mu$ related to the mass parameter.

A regular black-hole solution can be obtained by choosing the characteristic function such that it is monotonically increasing for the relevant range of $\psi$ and diverges at a finite value $\psi=\psi_0$. In this case, the metric function approaches
\begin{equation}
f(r)=1-r^2\psi_0+\mathcal{O}(r^{D-1})
\end{equation}
near the origin, corresponding to a de Sitter-like core rather than a curvature singularity.

For the Dymnikova spacetime, the characteristic function can be chosen as
\begin{equation}
h(\psi)=\frac{\psi}{
1+\alpha\psi
W_0\left(
-\frac{e^{-1/(\alpha\psi)}}{\alpha\psi}
\right)
},
\qquad
\alpha>0,
\end{equation}
where $W_0$ denotes the principal branch of the Lambert $W$ function. This yields
\begin{equation}
\psi(r)=\frac{\mu}{r^{D-1}}
\left(
1-e^{-r^{D-1}/(\alpha\mu)}
\right),
\end{equation}
and hence
\begin{equation}
f(r)=1-\frac{\mu}{r^{D-3}}\left(1-e^{-r^{D-1}/(\alpha\mu)}\right).
\end{equation}
In four dimensions, identifying $\mu=2M$ and $\alpha=\ell^2$, this reduces to
\begin{equation}
f(r)=1-\frac{2M}{r}\left(1-e^{-r^3/(2M\ell^2)}\right),
\label{eq:gmetric}
\end{equation}
which is the standard Dymnikova metric.

It is important to note that this construction is intrinsically nonperturbative in the higher-curvature corrections. In particular, the characteristic function containing the Lambert $W_0$ function cannot be represented by a finite polynomial in $\psi$. Therefore, the Dymnikova solution cannot be obtained by truncating the higher-curvature expansion at any finite order. Rather, it requires the full nonperturbative structure associated with the infinite tower of higher-curvature corrections~\cite{Konoplya:2024dymnikova}.

More recently, Borissova and Carballo-Rubio further developed this idea by constructing four-dimensional nonpolynomial quasitopological gravities in which regular black-hole spacetimes arise as vacuum solutions~\cite{Borissova:2026regular}. Their construction is based on nonpolynomial gravitational actions whose equations of motion remain second order on static and spherically symmetric backgrounds. In this framework, the metric function is determined by an algebraic equation involving theory-dependent characteristic functions. In particular, they showed that the spacetime described by the following static and spherically symmetric metric is the Dymnikova spacetime:
\begin{equation}
ds^2=-\left(1-\frac{2M}{r}\left(1-e^{-r^3/(2M\ell^2)}\right)\right)dt^2+\dfrac{dr^2}{\left(1-\frac{2M}{r}\left(1-e^{-r^3/(2M\ell^2)}\right)\right)}+r^2(d\theta^2+\sin^2\theta d\phi^2).
\label{metric Dymnikova}
\end{equation}
They further demonstrated that this spacetime arises as a vacuum solution of a four-dimensional nonpolynomial quasitopological gravity, whose corresponding characteristic function can be expressed in terms of the principal branch of the Lambert $W$ function. This provides a purely gravitational interpretation of the Dymnikova spacetime in four dimensions, without the need to introduce an explicit matter source.

These developments provide two complementary perspectives on the gravitational origin of the Dymnikova spacetime. The construction of Konoplya and Zhidenko embeds the Dymnikova geometry into a framework involving an infinite tower of higher-curvature corrections, emphasizing its intrinsically nonperturbative character. The subsequent construction by Borissova and Carballo-Rubio demonstrates that the same geometry can be realized explicitly as a vacuum solution of a four-dimensional nonpolynomial quasitopological gravity. This provides a particularly useful theoretical framework for interpreting the Dymnikova spacetime as a purely gravitational regular black-hole solution in four dimensions.

The existence of an event horizon is not guaranteed for arbitrary values of the parameter $\ell$. To clarify the parameter range in which the Dymnikova spacetime represents a black hole, we consider the horizon condition
\begin{align}
f(r)=0,
\end{align}
where $f(r)$ is given by Eq~(\ref{eq:gmetric}).

Introducing the dimensionless variables
\begin{align}
x=\frac{r}{2M},
\qquad
\lambda=\frac{\ell}{2M},
\end{align}
the horizon equation can be written as
\begin{align}
1-\frac{1}{x}
\left(
1-e^{-x^3/\lambda^2}
\right)=0.
\end{align}
Equivalently,
\begin{align}
x=1-e^{-x^3/\lambda^2}.
\end{align}
For a given value of $\ell$, the number of positive roots of this equation determines the causal structure of the spacetime.

For sufficiently small $\ell$, the metric function possesses two positive zeros, corresponding to an inner and an outer horizon. As $\ell$ increases, these two horizons approach each other and eventually merge at a critical value $\ell=\ell_{\rm crit}$. The critical configuration is characterized by a degenerate horizon satisfying
\begin{equation}
f(r_{\rm crit})=0,
\qquad
f'(r_{\rm crit})=0.
\end{equation}
These conditions determine the critical values of the horizon radius and the parameter $\ell$. Solving them gives
\begin{equation}
r_{\rm crit}=2M x_{\rm crit},
\end{equation}
where $x_{\rm crit}$ satisfies
\begin{align}
\lambda^2_{\rm crit}-3x_{\rm crit}^2(1-x_{\rm crit})=0.
\end{align}
The corresponding critical value of $\ell$ is
\begin{align}
\ell_{\rm crit}=2M\sqrt{3x_{\rm crit}^2(1-x_{\rm crit})}\simeq 1.13792M,
\end{align}
where
\begin{align}
x_{\rm crit}\simeq 0.851001.
\end{align}

Thus, the Dymnikova spacetime describes a black hole only for
\begin{align}
\ell<\ell_{\rm crit},
\end{align}
whereas at $\ell=\ell_{\rm crit}$ the inner and outer horizons coincide and the spacetime is extremal. For
\begin{align}
\ell>\ell_{\rm crit},
\end{align}
no horizon exists.

The critical value therefore provides an upper bound on the parameter $\ell$ for the Dymnikova black-hole geometry. In the following analysis, we restrict the parameter range to $\ell<\ell_{\rm crit}$ so that the background represents a genuine black hole with an outer event horizon.

\subsection{Photon Sphere and Critical Impact Parameter}

We next review the conditions for the existence of a photon sphere and the corresponding critical impact parameter in the Dymnikova spacetime. In the four-dimensional case, the metric is written as Eq~(\ref{metric Dymnikova}).

For null geodesics in the equatorial plane, the impact parameter $u$ is given by
\begin{align}
u(r)&=\sqrt{\frac{r^2}{f(r)}}.
\end{align}
The circular null orbit, or photon sphere, is determined by the extremum of the effective potential for null geodesics, which is proportional to $f(r)/r^2$. Thus, the photon-sphere radius $r_{\rm ph}$ is determined by
\begin{align}
\frac{d}{dr}\left(\frac{f(r)}{r^2}\right)=0,
\end{align}
which is equivalently written as
\begin{align}
r f'(r)-2f(r)=0.
\label{eq:photon_sphere_condition}
\end{align}

It is useful to introduce the dimensionless variables
\begin{align}
z=\frac{r^3}{2M\ell^2}.
\end{align}
The photon-sphere condition can then be written as
\begin{align}
r_{\rm ph}&=3M(1-(1+z)e^{-z}),\\
\ell&=\sqrt{\frac{27}{2}}M\frac{(1-(1+z)e^{-z})^{3/2}}{\sqrt{z}}.
\label{eq:photon_sphere_dimensionless}
\end{align}

The existence of a photon sphere is determined by the range of $\ell$ for which the above equations admit positive solutions for $z$. The critical value of $\ell$ is attained when $z$ takes the value
\begin{align}
z_{\rm ph,crit}\simeq3.94343M,
\end{align}
and is given by
\begin{align}
\ell_{\rm ph,crit}\simeq1.59063M.
\end{align}
At this critical value, two circular null orbits merge. The corresponding photon-sphere radius is
\begin{align}
r_{\rm ph,crit}\simeq2.71257M.
\end{align}

This critical value should be distinguished from the critical value associated with the existence of horizons. The Dymnikova spacetime possesses horizons only for
\begin{align}
\ell
&<
\ell_{\rm H,crit},
\qquad
\ell_{\rm H,crit}
\simeq
1.13792,M.
\end{align}
Thus,
\begin{align}
\ell_{\rm H,crit}
&<
\ell_{\rm ph,crit}.
\end{align}
Therefore, throughout the parameter region in which the Dymnikova spacetime represents a black hole, an outer photon sphere is present. This outer photon sphere corresponds to the unstable circular null orbit relevant to the strong-deflection limit.

For a photon sphere located at $r=r_{\rm ph}$, the critical impact parameter is determined by
\begin{align}
u_c
&=
\frac{r_{\rm ph}}
{\sqrt{f(r_{\rm ph})}}.
\label{eq:bc_Dymnikova}
\end{align}
The critical impact parameter separates qualitatively different photon trajectories. Photons with $u>u_c$ are scattered back to the observer, whereas photons with $u<u_c$ are captured by the black hole. As $u$ approaches $u_c$ from above, the photon spends an increasingly long time in the vicinity of the unstable photon sphere, resulting in the logarithmic divergence of the deflection angle characteristic of the strong-deflection regime.

The photon sphere and the critical impact parameter therefore provide the essential quantities for the strong-deflection analysis of the Dymnikova spacetime. In the presence of EFT corrections, the photons propagate along null geodesics of the polarization-dependent effective optical metric. Consequently, the photon-sphere radius and the critical impact parameter are modified for the PPL and PPM modes.

In this work, we consider four representative values of $\ell$, denoted by $\ell=\sqrt{\frac{27}{2}}\frac{(1-101e^{-100})^{3/2}}{10},\sqrt{\frac{27}{2}}\frac{(1-11e^{-10})^{3/2}}{\sqrt{10}},$
$\sqrt{\frac{27}{2}}\frac{(1-6e^{-5})^{3/2}}{\sqrt{5}},\sqrt{\frac{27}{2}}\frac{(1-5e^{-4})^{3/2}}{2}$, and compare the resulting observables with those of the Schwarzschild spacetime. These values are chosen to cover different regimes of the Dymnikova spacetime: one close to the Schwarzschild limit, one near the critical value $\ell_{\rm ph,crit}$, and two intermediate cases between these regimes.

\section{Photon Propagation and Effective Geometry in EFT-Corrected Hayward black hole}
\label{sec:photon}

Gravitational lensing by compact objects provides a powerful probe of strong-field gravity and the underlying spacetime geometry~\cite{Einstein:1936llh,Virbhadra:1999nm,Bozza:2001xd,Bozza:2002zj,Gibbons:2008rj}. Beyond the geometry of the background spacetime, the propagation of light can also be modified by non-minimal couplings between the electromagnetic field and spacetime curvature. Such curvature couplings arise naturally from quantum corrections to electrodynamics in curved spacetime~\cite{Drummond:1979pp} and can lead to polarization-dependent photon propagation, giving rise to a gravitational analogue of birefringence~\cite{Chen:2015cpa,Lu:2016gsf}. Among the possible curvature couplings, the interaction between the electromagnetic field and the Weyl tensor has been extensively studied, including its effects on photon propagation and gravitational lensing in black-hole spacetimes~\cite{Drummond:1979pp,Chen:2015cpa,Lu:2016gsf,Chen:2016hil}.

In this work, we investigate a broader class of curvature-photon couplings, including both Weyl and Ricci tensor interactions, in the context of a Hayward regular black hole. We study how these EFT corrections modify photon propagation and whether they can lead to observable differences between the Hayward and Schwarzschild black holes through strong gravitational lensing. In contrast to the minimally coupled case, the curvature-photon interactions modify the photon propagation law and can lead to polarization-dependent photon trajectories. We derive the corresponding effective optical metrics from the photon dispersion relation in the eikonal approximation and use them to analyze the strong-deflection gravitational lensing in the Hayward spacetime. In particular, we determine the corrections to the photon-sphere radius and the logarithmically divergent part of the deflection angle, and investigate how these corrections depend on the Hayward parameter and the EFT couplings. This framework allows us to assess whether the combined effects of the nontrivial regular black-hole geometry and curvature-dependent photon propagation can leave characteristic signatures that distinguish a Hayward regular black hole from the Schwarzschild black hole.

\subsection{Effective Optical Metric for Photon Propagation in EFT-Corrected Dymnikova Spacetim}

The starting point of our analysis is a four-dimensional nonpolynomial quasitopological gravity coupled to the electromagnetic field, supplemented by effective field theory (EFT) corrections that modify photon propagation. We consider the effective action
\begin{align}
S=\int d^{4}x\sqrt{-g}\Biggl[
\mathcal{L}(R,R_{\mu\nu},C_{\mu\nu\rho\sigma})
-\frac{1}{4}F_{\mu\nu}F^{\mu\nu}
+\alpha R F_{\mu\nu}F^{\mu\nu}
+\beta R_{\mu\nu}F^{\mu\rho}F^{\nu}{}_{\rho}
+\gamma C^{\mu\nu\rho\sigma}F_{\mu\nu}F_{\rho\sigma}
\Biggr],
\label{eq:action}
\end{align}
where $\mathcal{L}$ is a nonpolynomial function of curvature invariants constructed from the Ricci scalar $R$, the Ricci tensor $R_{\mu\nu}$, and the Weyl tensor $C_{\mu\nu\rho\sigma}$, and contains the same action terms as those appearing in Eq.~\eqref{action 1}. Here, $F_{\mu\nu}$ denotes the electromagnetic field strength, and $\alpha$, $\beta$, and $\gamma$ are EFT coupling constants with dimensions of length squared.

We regard the curvature-photon interaction terms as the leading higher-derivative operators in the derivative expansion of the underlying effective theory. In this work, we restrict the analysis to the leading nontrivial order in this expansion and retain only operators containing four derivatives in total. Higher-derivative operators are therefore consistently neglected, as their effects are suppressed by additional powers of the EFT scale. The corresponding corrections are treated perturbatively, and only terms at first order in the EFT coupling constants are retained throughout the analysis.

The first term in the action describes the gravitational sector through a four-dimensional nonpolynomial quasitopological gravity. The Lagrangian $\mathcal{L}$ is a nonpolynomial function of curvature invariants constructed from the Ricci scalar $R$, the Ricci tensor $R_{\mu\nu}$, and the Weyl tensor $C_{\mu\nu\rho\sigma}$. This theory admits the Dymnikova spacetime as a vacuum solution. The electromagnetic field is minimally coupled to this gravitational background through the standard Maxwell term. The remaining terms describe non-minimal interactions between the electromagnetic field and spacetime curvature. In particular, the $\beta$ and $\gamma$ terms couple the electromagnetic field to the Ricci and Weyl tensors, respectively, and modify the photon propagation law in the curved spacetime. The $\alpha$ term involving the Ricci scalar is also included for completeness.

We do not introduce additional matter fields or direct couplings between the electromagnetic field and independent scalar fields. Instead, the effects of the underlying gravitational theory are encoded in the background geometry through the Dymnikova solution. In the exterior region of the black hole, where the curvature becomes weaker away from the horizon, higher-order curvature contributions are suppressed. We therefore focus on the leading curvature-dependent corrections to photon propagation. In particular, the curvature-photon interactions $R F_{\mu\nu}F^{\mu\nu}$, $R_{\mu\nu}F^{\mu\rho}F^{\nu}{}_{\rho}$, and $C^{\mu\nu\rho\sigma}F_{\mu\nu}F_{\rho\sigma}$ provide the leading non-minimal couplings between the electromagnetic field and the background curvature. These terms capture the leading effects of the curved background on photon propagation while consistently neglecting higher-order EFT operators.

Since the background electromagnetic field is assumed to vanish, $F_{\mu\nu}=0$, the curvature-photon interaction terms do not modify the background geometry at first order in the EFT couplings. Nevertheless, they contribute directly to the photon equations of motion and modify the photon dispersion relation at leading order in the eikonal approximation. These curvature-dependent interactions are therefore the relevant EFT operators for studying the propagation of photons and their gravitational lensing in the background spacetime considered below.

We take the Dymnikova regular black hole as the background spacetime. Unlike the Schwarzschild spacetime, the Dymnikova geometry is generally non-Ricci-flat, and its curvature structure is determined by the parameter $\ell$. Consequently, both Ricci- and Weyl-curvature contributions can affect photon propagation through the EFT interactions. In particular, the $\beta$ term probes the Ricci curvature of the Dymnikova background, while the $\gamma$ term couples the electromagnetic field to the Weyl curvature and therefore captures the effect of the gravitational curvature on photon propagation. The $\alpha$ term provides an additional coupling to the Ricci scalar. This provides a natural framework for investigating how curvature-dependent modifications of photon propagation affect gravitational lensing in the Dymnikova spacetime and how they differ from the corresponding Schwarzschild case.

Taking the variation of the action~\eqref{eq:action} with respect to the electromagnetic field yields the modified Maxwell equation,
\begin{align}
\nabla_{\mu}\left(F^{\mu\nu}-4\alpha R F^{\mu\nu}-2\beta R_{\mu\rho}F^{\rho\nu}-2\beta R^{\nu\rho}F^{\mu}{}_{\rho}-4\gamma C^{\mu\nu\rho\sigma}F_{\rho\sigma}\right)=0 .
\label{eq:maxwell-mod}
\end{align}
To study photon propagation in the geometric-optics regime, we employ the eikonal approximation and write the electromagnetic potential in terms of a rapidly varying phase and a slowly varying amplitude. Denoting the wave vector by $k_\mu=\nabla_\mu\theta$ and the polarization vector by $a^\mu$, we impose the transversality condition $k_\mu a^\mu=0$. Substituting the eikonal ansatz into the modified Maxwell equation and retaining the leading-order terms in the eikonal expansion, we obtain the following equation governing photon propagation:
\begin{align}
(1-4\alpha R)k_{\mu}k^{\mu}a^{\nu}-2\beta R_{\mu\rho}k^\mu k^\rho a^\nu+2\beta R_{\mu\rho}k^\mu a^\rho k^\nu-2\beta R^{\rho\nu}a_\rho k_{\mu}k^{\mu}+8\gamma C^{\mu\nu\rho\sigma}k_{\sigma}k_{\mu}a_{\rho}=0 .
\label{eq:eom-photon}
\end{align}
Since we treat the EFT corrections perturbatively, the zeroth-order propagation is governed by the GR null condition $k_\mu k^\mu=0$. Consequently, the terms proportional to $k_\mu k^\mu$, namely the term proportional to $\alpha R$ and the fourth term in Eq.~\eqref{eq:eom-photon}, do not contribute to the dispersion relation at first order in the EFT couplings. We therefore omit these terms in the following analysis.

\subsection{Dymnikova Spacetime and the Weyl Tensor in an Orthonormal Frame}

The metric of the Dymnikova spacetime \cite{Dymnikova:1992ux} can be written as Eq~(\ref{metric Dymnikova}). Building on the treatment developed in~\cite{Daniels:1993yi,Shore:1995fz,Daniels:1995yw,Chen:2016hil,Kanai:2026ohg,Kanai:2026bag}, we employ an orthonormal vierbein $e^{a}_{\ \mu}$ satisfying
\begin{align}
g_{\mu\nu}=\eta_{ab}e^{a}_{\ \mu}e^{b}_{\ \nu},
\end{align}
and define the antisymmetric tensor

\begin{align}
U^{ab}_{\ \ \mu\nu}
=e^{a}_{\ \mu}e^{b}_{\ \nu}-e^{a}_{\ \nu}e^{b}_{\ \mu}.
\end{align}

With respect to this orthonormal basis, the Weyl tensor admits a convenient bilinear decomposition in terms of the vierbein combinations above:

\begin{align}
C_{\mu\nu\rho\sigma}
=\mathcal{A}\Big(
U^{01}_{\mu\nu}U^{01}_{\rho\sigma}
-U^{23}_{\mu\nu}U^{23}_{\rho\sigma}\Big)
+\mathcal{B}\Big(
U^{02}_{\mu\nu}U^{02}_{\rho\sigma}
+U^{03}_{\mu\nu}U^{03}_{\rho\sigma}
-U^{12}_{\mu\nu}U^{12}_{\rho\sigma}
-U^{13}_{\mu\nu}U^{13}_{\rho\sigma}
\Big),
\label{eq:weyl-general}
\end{align}

where $\mathcal{A}$ and $\mathcal{B}$ are scalar functions determined by the radial coordinate $r$, the metric function $f(r)$, and their derivatives. For the spacetime considered here, these functions take the form

\begin{align}
\mathcal{A}
&=\frac{e^{-\frac{r^3}{2M\ell^2}}\left(-4M^2\ell^4\left(-1+e^{\frac{r^3}{2M\ell^2}}\right)-2M\ell^2 r^3+3r^6\right)}{2\ell^4 r^3},
\label{A}\\
\mathcal{B}
&=\frac{e^{-\frac{r^3}{2M\ell^2}}\left(8M^2\ell^4\left(-1+e^{\frac{r^3}{2M\ell^2}}\right)-20M\ell^2 r^3+3r^6\right)}{8\ell^4 r^3}.
\label{B}
\end{align}

This decomposition provides a useful representation of the Weyl tensor for a static, spherically symmetric spacetime and serves as the starting point for deriving the modified photon propagation equation induced by the Weyl-photon coupling.

\subsection{Polarization decomposition and the light-cone condition}

To formulate the photon equation of motion~\eqref{eq:eom-photon} for the Dymnikova spacetime, it is useful to define the momentum-contracted quantities~\cite{Daniels:1993yi,Shore:1995fz,Daniels:1995yw}
\begin{equation}
l_{\nu}=k^{\mu}U^{01}_{\ \ \mu\nu},\qquad
n_{\nu}=k^{\mu}U^{02}_{\ \ \mu\nu},\qquad
m_{\nu}=k^{\mu}U^{23}_{\ \ \mu\nu},
\label{eq:lnr}
\end{equation}
together with the related combinations
\begin{equation}
p_{\nu}=k^{\mu}U^{12}_{\ \ \mu\nu},\qquad
q_{\nu}=k^{\mu}U^{03}_{\ \ \mu\nu},\qquad
r_{\nu}=k^{\mu}U^{13}_{\ \ \mu\nu},
\end{equation}
all of which are orthogonal to $k^\nu$. Contracting Eq.~\eqref{eq:eom-photon} with the appropriate independent polarization vectors reduces the photon equation of motion to a homogeneous linear system for the three independent polarization amplitudes $a\cdot l$, $a\cdot n$, and $a\cdot m$,
\begin{equation}
\begin{pmatrix}
K_{11} & 0 & 0\\
K_{21} & K_{22} & K_{23}\\
0 & 0 & K_{33}
\end{pmatrix}
\begin{pmatrix}
a\cdot l\\
a\cdot n\\
a\cdot m
\end{pmatrix}
=0.
\label{eq:Kmatrix}
\end{equation}
The coefficients $K_{ij}$ depend on the metric functions, the curvature tensors, and the components of the wave vector. At the order relevant to the present perturbative analysis, they are given by
\begin{align}
K_{11}={}&
(1+8\gamma\mathcal{A})
(g_{00}k^0k^0+g_{11}k^1k^1)
+(1+8\gamma\mathcal{B})
(g_{22}k^2k^2+g_{33}k^3k^3)
-2\beta R_{\mu\nu}k^\mu k^\nu,
\\
K_{21}={}&
8\gamma(\mathcal{A}-\mathcal{B})
\sqrt{g_{11}g_{22}}\,k^1k^2,
\\
K_{22}={}&
(1+8\gamma\mathcal{B})
(g_{00}k^0k^0+g_{11}k^1k^1
+g_{22}k^2k^2+g_{33}k^3k^3)
-2\beta R_{\mu\nu}k^\mu k^\nu,
\\
K_{23}={}&
-8\gamma(\mathcal{A}-\mathcal{B})
\sqrt{-g_{00}g_{33}}\,k^0k^3,
\\
K_{33}={}&
(1+8\gamma\mathcal{B})
(g_{00}k^0k^0+g_{11}k^1k^1)
+(1+8\gamma\mathcal{A})
(g_{22}k^2k^2+g_{33}k^3k^3)
-2\beta R_{\mu\nu}k^\mu k^\nu.
\end{align}

For a nontrivial solution, the determinant of the coefficient matrix must vanish. At the order considered here, the determinant factorizes into three propagation conditions,
\begin{align}
\det K=\tilde K_{11}\tilde K_{22}\tilde K_{33}=0.
\label{eq:Kdet-eq}
\end{align}
The first condition, $\tilde K_{11}=0$, gives
\begin{align}
&\left[(1+8\gamma\mathcal{A})g_{00}-2\beta R_{00}\right]k^0k^0
+\left[(1+8\gamma\mathcal{A})g_{11}-2\beta R_{11}\right]k^1k^1
\notag\\
&\quad
+\left[(1+8\gamma\mathcal{B})g_{22}-2\beta R_{22}\right]k^2k^2
+\left[(1+8\gamma\mathcal{B})g_{33}-2\beta R_{33}\right]k^3k^3
=0.
\label{eq:PPL}
\end{align}
This mode corresponds to a polarization vector lying within the orbital plane. Following Refs.~\cite{Chen:2015cpa,Lu:2016gsf}, we refer to this polarization as the \emph{PPL} mode.

The second condition, $\tilde K_{22}=0$, reduces to the standard null condition
\begin{equation}
g_{\mu\nu}k^\mu k^\nu=0.
\end{equation}
It corresponds to the non-propagating polarization branch in the curvature-coupled system and does not represent an independent physical polarization mode. We therefore discard this branch, following the treatment in Refs.~\cite{Drummond:1979pp,Chen:2015cpa,Lu:2016gsf}.

The third condition, $\tilde K_{33}=0$, is given by
\begin{align}
&\left[(1+8\gamma\mathcal{B})g_{00}-2\beta R_{00}\right]k^0k^0
+\left[(1+8\gamma\mathcal{B})g_{11}-2\beta R_{11}\right]k^1k^1
\notag\\
&\quad
+\left[(1+8\gamma\mathcal{A})g_{22}-2\beta R_{22}\right]k^2k^2
+\left[(1+8\gamma\mathcal{A})g_{33}-2\beta R_{33}\right]k^3k^3
=0.
\label{eq:PPM}
\end{align}
The corresponding polarization vector is directed perpendicular to the orbital plane, and we refer to this branch as the \emph{PPM} mode, following Ref.~\cite{Chen:2016hil}.

Consequently, Eqs.~\eqref{eq:PPL} and~\eqref{eq:PPM} characterize the two physical propagation branches associated with the polarization of curvature-coupled photons in the equatorial plane of a general static and spherically symmetric spacetime. Each branch determines a different effective light cone, or equivalently, a distinct effective optical metric. Polarization-dependent photon trajectories therefore emerge, providing a gravitational counterpart of birefringence. When the curvature couplings are switched off, $\beta,\gamma\rightarrow0$, both propagation equations recover the usual null condition associated with the background spacetime, so that the two polarization states become degenerate.

The curvature-photon couplings introduced above modify the light-cone structure of photon propagation in the Dymnikova spacetime. For the analysis that follows, it is useful to recast the two physical propagation conditions associated with the PPL and PPM polarization modes in terms of effective optical metrics. Working to first order in the EFT coupling constants, the two polarization-dependent effective metrics can be expressed in a unified form as
\begin{align} 
\label{metric_1} 
ds^2=&-f(r)\left(1+\beta\mathcal{C}\right)^{-1}dt^2+\frac{1}{f(r)}\left(1+\beta\mathcal{C}\right)^{-1}dr^2+r^2\left(1+\beta\mathcal{D}\right)^{-1}\left(\dfrac{1+8\gamma\mathcal{A}}{1+8\gamma\mathcal{B}}\right)^s(d\theta^2+\sin^2\theta d\phi^2),\\ 
\mathcal{C}=&-\dfrac{3re^{-\frac{r^3}{2M\ell^2}}(4M\ell^2-3r^3)}{2M\ell^4},\\ 
\mathcal{D}=&-\frac{6e^{-\frac{r^3}{2M\ell^2}}}{\ell^2},
\end{align}
where $f(r)$, $\mathcal{A}$ and $\mathcal{B}$ are defined in Eqs.~\eqref{eq:gmetric}, (\ref{A}) and (\ref{B}), respectively, and $s=+1$ and $s=-1$ denote the PPL and PPM polarization modes, respectively. The functions $\mathcal{A}$ and $\mathcal{B}$ characterize the contribution from the Weyl curvature, whereas $\mathcal{C}$ and $\mathcal{D}$ originate from the coupling to the Ricci curvature. The dependence of the effective metric on the polarization mode therefore arises from the interaction between the photon and the Weyl curvature.

In the limit $\beta,\gamma\rightarrow0$, the effective metric reduces to the underlying Dymnikova spacetime. We regard the EFT corrections as perturbative contributions and consistently retain only terms linear in the EFT coupling constants. The effective optical metrics in Eq.~\eqref{metric_1} thus serve as the starting point for investigating photon trajectories and strong-deflection gravitational lensing in the EFT-corrected Dymnikova regular black hole spacetime.

Having established the polarization-dependent effective optical metrics for the EFT-corrected Dymnikova spacetime, we have obtained the framework necessary to study its strong-deflection lensing properties. In the next section, we apply these effective metrics to derive the strong-deflection angle and the corresponding coefficients for the PPL and PPM modes, and investigate the effects of the EFT corrections in comparison with the Schwarzschild spacetime. We then turn to the time delay between different photon trajectories in the subsequent section, focusing on its dependence on the polarization modes and EFT corrections. Together, these analyses provide a basis for examining the observational differences between the EFT-corrected Dymnikova and Schwarzschild spacetimes through strong-deflection lensing observables.

\section{Strong-Deflection Limit in EFT-Corrected Dymnikova black hole}
\label{sec:strong lens}

In the previous section, we investigated the strong-deflection limit of light propagation in the Hayward black hole spacetime in the absence of EFT corrections, providing the baseline lensing observables for comparison. In this section, we incorporate EFT corrections to photon propagation in the Hayward background. These curvature-dependent corrections modify the photon propagation law and lead to polarization-dependent effective optical metrics for the physical photon modes. We investigate how these modifications affect the photon-sphere radius, critical impact parameter, and strong-deflection observables, and assess whether the combined effects of the Hayward geometry and EFT-corrected photon propagation can provide characteristic signatures for distinguishing the Hayward regular black hole from the Schwarzschild black hole.

We briefly review the strong-deflection formalism for a general static and spherically symmetric spacetime. Restricting the photon motion to the equatorial plane, $\theta=\pi/2$, the metric is written as
\begin{align}
ds^2=-A(r)dt^2+B(r)dr^2+C(r)d\phi^2,
\end{align}
where $A(r)$, $B(r)$, and $C(r)$ are functions of the radial coordinate. We denote by $r_0$ the radius of closest approach and define $A_0=A(r_0)$ and $C_0=C(r_0)$. Following Bozza~\cite{Bozza:2002zj,Bozza:2002af,Bozza:2003cp}, we introduce
\begin{align}
\label{zcoor}
z=1-\frac{r_0}{r}.
\end{align}

The total change in the azimuthal angle is
\begin{align}
\Delta\phi\equiv I(r_0)=\int_0^1 R(z,r_0)f(z,r_0),dz,
\end{align}
where
\begin{align}
R(z,r_0)&=\frac{2r_0\sqrt{A(r(z))B(r(z))C_0}}{C(r(z))(1-z)^2},\\
f(z,r_0)&=\frac{1}{\sqrt{A_0-A(r(z))C_0/C(r(z))}}.
\end{align}
Here and below, a prime denotes differentiation with respect to $r$, and quantities carrying the subscript $0$ are evaluated at $r=r_0$.

The divergence of the integral originates from $f(z,r_0)$ as $z\rightarrow0$. Expanding its argument around $z=0$ gives
\begin{align}
f(z,r_0)\simeq f_0(z,r_0)=\frac{1}{\sqrt{p(r_0)z+q(r_0)z^2}},
\label{pole expansion}
\end{align}
where
\begin{align}
p(r_0)
&=\frac{r_0}{C_0}\left(C_0'A_0-C_0A_0'\right),
\end{align}
and $q(r_0)$ is determined by the corresponding second-order expansion. The divergence occurs when $p(r_0)=0$, which determines the radius of the unstable circular photon orbit, or photon sphere,
\begin{align}
\left.
\frac{d}{dr}\left(\frac{C(r)}{A(r)}\right)\right|_{r=r_{\rm ph}}=0.
\label{ph radius}
\end{align}
The corresponding critical impact parameter is
\begin{align}
u_{\rm ph}
=\sqrt{\frac{C_{\rm ph}}{A_{\rm ph}}},
\label{critical impact}
\end{align}
where $A_{\rm ph}=A(r_{\rm ph})$ and $C_{\rm ph}=C(r_{\rm ph})$.

We decompose the integral into divergent and regular parts,
\begin{align}
I(r_0)=I_D(r_0)+I_R(r_0),
\end{align}
where
\begin{align}
I_D(r_0)&=\int_0^1R(0,r_{\rm ph})f_0(z,r_0),dz,\\
I_R(r_0)&=\int_0^1g(z,r_0),dz,
\label{regular angle}
\end{align}
with
\begin{align}
g(z,r_0)=R(z,r_0)f(z,r_0)-R(0,r_{\rm ph})f_0(z,r_0).
\end{align}
The divergent contribution is
\begin{align}
I_D(r_0)=\frac{2R(0,r_{\rm ph})}{\sqrt{q(r_0)}}\log\left[\frac{\sqrt{q(r_0)}+\sqrt{p(r_0)+q(r_0)}}{\sqrt{p(r_0)}}\right].
\end{align}
Near the photon sphere,
\begin{align}
I_D(r_0)=-a\log\left(\frac{r_0}{r_{\rm ph}}-1\right)+b_D+\mathcal{O}\left((r_0-r_{\rm ph})\log(r_0-r_{\rm ph})\right),
\end{align}
where
\begin{align}
a&=\frac{R(0,r_{\rm ph})}{\sqrt{q(r_{\rm ph})}},\\
b_D&=\frac{R(0,r_{\rm ph})}{\sqrt{q(r_{\rm ph})}}\log2.
\end{align}

The regular contribution is
\begin{align}
I_R(r_0)=\int_0^1g(z,r_{\rm ph}),dz+\mathcal{O}(r_0-r_{\rm ph}),
\end{align}
with
\begin{align}
b_R=I_R(r_{\rm ph})=\int_0^1g(z,r_{\rm ph}),dz.
\end{align}
Thus, the total deflection angle is
\begin{align}
I(r_0)=-a\log\left(\frac{r_0}{r_{\rm ph}}-1\right)+b+\mathcal{O}\left((r_0-r_{\rm ph})\log(r_0-r_{\rm ph})\right),
\end{align}
where
\begin{align}
b=-\pi+b_D+b_R.
\end{align}

Introducing the impact parameter
\begin{align}
u=\sqrt{\frac{C_0}{A_0}}.
\label{impact para}
\end{align}
We have $u=u_{\rm ph}$ at the photon sphere. Its expansion around the critical orbit is
\begin{align}
\label{c_def}
u-u_{\rm ph}=c(r_0-r_{\rm ph})^2
+\mathcal{O}\left((r_0-r_{\rm ph})^3\right),
\end{align}
where $c$ is determined by the metric functions and their derivatives evaluated at the photon sphere. Using $u\simeq D_{OL}\theta$, the strong-deflection angle becomes
\begin{align}
\alpha(\theta)=-\bar a\log\left(\frac{\theta D_{OL}}{u_{\rm ph}}-1\right)+\bar b,
\end{align}
where
\begin{align}
\bar a&=\frac{a}{2}=\frac{R(0,r_{\rm ph})}{2\sqrt{q(r_{\rm ph})}},\\
\bar b&=b+\frac{a}{2}\log\left(\frac{cr_{\rm ph}^2}{u_{\rm ph}}\right)\notag\\
&=-\pi+b_R+\bar a\log\left(\frac{2q(r_{\rm ph})}{A_{\rm ph}}\right).
\end{align}
The coefficient $\bar a$ determines the strength of the logarithmic divergence, while $\bar b$ gives the constant term in the leading strong-deflection expansion. Together with $u_{\rm ph}$, they characterize the leading strong-deflection behavior and the corresponding relativistic lensing observables. In the following subsection, we apply this formalism to the Dymnikova spacetime including the EFT corrections to photon propagation, and investigate whether the Dymnikova black hole can be distinguished from the Schwarzschild black hole through the resulting lensing observables.

\subsection{Strong Deflection Analysis of EFT-corrected Dymnikova black hole}

In this subsection, we investigate gravitational lensing in the strong-deflection regime, taking into account the effects of the curvature-photon couplings introduced in Eq.~\eqref{eq:action}. We consider the Dymnikova regular black hole as the background spacetime and focus on the modifications to photon propagation induced by the EFT corrections. Following the convention adopted in the previous section, we set $M=1$ and express all quantities in dimensionless form by measuring lengths in units of the black-hole mass.

For photons propagating in the equatorial plane, the EFT-corrected propagation law derived from the curvature-photon interactions can be described in terms of the effective optical metrics obtained in the previous section. The two physical polarization modes, PPL and PPM, experience different effective geometries as a consequence of their couplings to the Weyl and Ricci curvatures. After setting $M=1$, the effective optical metrics in Eq.~\eqref{metric_1} can be rewritten in the following form:
\begin{align}
\label{metric_2}
ds^2=&-f(r)\left(1+\beta\mathcal{C}\right)^{-1}dt^2+\frac{1}{f(r)}\left(1+\beta\mathcal{C}\right)^{-1}dr^2+r^2\left(1+\beta\mathcal{D}\right)^{-1}\left(\dfrac{1+8\gamma\mathcal{A}}{1+8\gamma\mathcal{B}}\right)^s(d\theta^2+\sin^2\theta d\phi^2),\\ 
\mathcal{A}=&\frac{e^{-\frac{r^3}{2\ell^2}}\left(-4\ell^4\left(-1+e^{\frac{r^3}{2\ell^2}}\right)-2\ell^2 r^3+3r^6\right)}{2\ell^4 r^3},\\
\mathcal{B}=&\frac{e^{-\frac{r^3}{2\ell^2}}\left(8\ell^4\left(-1+e^{\frac{r^3}{2\ell^2}}\right)-20\ell^2 r^3+3r^6\right)}{8\ell^4 r^3}.\\
\mathcal{C}=&-\dfrac{3re^{-\frac{r^3}{2\ell^2}}(4\ell^2-3r^3)}{2\ell^4},\\ 
\mathcal{D}=&-\frac{6e^{-\frac{r^3}{2\ell^2}}}{\ell^2}.
\end{align}
In this paper, we perform our analysis using dimensionless coordinates, which correspond to setting $M=1$. In these coordinates, the metric above takes the corresponding dimensionless form.

The strong-deflection analysis then allows us to derive the lensing observables while retaining the dependence on the physical parameters. In particular, the deflection angle is decomposed into a logarithmically divergent part and a regular contribution. The divergent part can be obtained analytically. For $\ell\rightarrow0$, corresponding to the Schwarzschild limit, the regular contribution $b_R$ can also be evaluated analytically. For the nonzero values of $\ell$ considered here, however, the integral defining $b_R$ cannot be evaluated in closed form. We therefore evaluate $b_R$ numerically for the corresponding parameter choices.

For the spacetime under consideration, the photon-sphere radius, including the leading EFT corrections, is given for each value of $\ell$ in Table~\ref{tab:photon_sphere}. Here $r_{\rm ph}^{(0)}$ denotes the radius of photon sphere in the Dymnikova spacetime.
\begin{table}[t]
\centering
\caption{Photon-sphere radius $r_{\rm ph}$ for different values of $\ell$, with $M=1$.}
\label{tab:photon_sphere}
\begin{tabular}{c|c|c|c}
\hline\hline
$\ell$ & Numerical value of $\ell$ & Numerical value of $r_{\rm ph}^{(0)}$ & $r_{\rm ph}$ \\
\hline
$0$ & $0$ & 3 & $3-\frac{4s\gamma}{3}$ \\
\hline
$\sqrt{\frac{27}{2}}\frac{(1-101e^{-100})^{3/2}}{10}$ & 0.36742 & 3.00000.& $3(1-101e^{-100})-\frac{4e^{100}(-301+e^{100})(-742500\beta+(-1495101+e^{100})s\gamma)}{3(-30101+e^{100})(-101+e^{100})^2}$ \\
\hline
$\sqrt{\frac{27}{2}}\frac{(1-11e^{-10})^{3/2}}{\sqrt{10}}$ & 1.16103 & 2.99850 & $3(1-11e^{-10})-\frac{4e^{10}(-31+e^{10})(-675\beta+(-1461+e^{10})s\gamma)}{3(-311+e^{10})(-11+e^{10})^2}$ \\
\hline
$\sqrt{\frac{27}{2}}\frac{(1-6e^{-5})^{3/2}}{\sqrt{5}}$ & 1.54454 & 2.87872& $3(1-6e^{-5})-\frac{4e^5(-16+e^5)(-75\beta+(-181+e^5)s\gamma)}{3(-81+e^5)(-6+e^5)^2}$ \\
\hline
$\sqrt{\frac{27}{2}}\frac{(1-5e^{-4})^{3/2}}{2}$ & 1.59063 & 2.72527 & $3(1-5e^{-4})-\frac{4e^4(-13+e^4)(-36\beta+(-93+e^4)s\gamma)}{3(-53+e^4)(-5+e^4)^2}$ \\
\hline\hline
\end{tabular}
\end{table}
The photon-sphere radius plays a central role in determining the properties of strong gravitational lensing. In particular, it determines the critical impact parameter $u_c$, which characterizes the boundary between photon trajectories that are scattered back to the observer and those that are captured by the central object. The critical impact parameter is directly related to the angular positions of relativistic images, providing a natural connection between the photon-sphere structure and observable lensing quantities. We therefore use the photon-sphere radius obtained above as a key ingredient in the subsequent analysis of the strong-deflection limit.

For the nonzero values of $\ell$ considered below, the integral defining the regular contribution $b_R$ does not admit a closed-form expression. Nevertheless, the strong-deflection coefficient $\bar a$, and the critical impact parameter $u_{\rm ph}$ can be obtained analytically for each value of $\ell$. We therefore evaluate only the regular contribution $b_R$ numerically for the nonzero values of $\ell$, while $\bar a$, and $u_{\rm ph}$ are determined analytically. The resulting expressions for $\bar a$, and $u_{\rm ph}$ are summarized in Tables~\ref{tab:bara}-\ref{tab:uph}.
 
\begin{table}[htbp]
\centering
\caption{Strong-deflection coefficient $\bar{a}$ for different values of $\ell$, including the leading-order EFT corrections, with $M=1$.}
\label{tab:bara}
\begin{tabular}{c|c}
\hline
$\ell$ & $\bar{a}$ \\
\hline
$0$ & $1+\dfrac{4s\gamma}{9}$\\
\hline
$\sqrt{\frac{27}{2}}\frac{(1-101e^{-100})^{3/2}}{10}$ & $\sqrt{\dfrac{-101+e^{100}}{-30101+e^{100}}}\left(1+4e^{200}\frac{-30000(-114387299+33698e^{100}+3601e^{200})\beta+(6770803789699-1991609e^{100}-213180303e^{200}+e^{300})s\gamma}{9(-30101+e^{100})^2(-101+e^{100})^3}\right)$ \\
\hline
 $\sqrt{\frac{27}{2}}\frac{(1-11e^{-10})^{3/2}}{\sqrt{10}}$ & $\sqrt{\dfrac{-11+e^{10}}{-311+e^{10}}}\left(1+4e^{20}\frac{-150(-26458+661e^{10}+47e^{20})\beta+(6761869-165537e^{10}-12333e^{20}+e^{30})s\gamma}{9(-311+e^{10})^2(-11+e^{10})^3}\right)$ \\
\hline
 $\sqrt{\frac{27}{2}}\frac{(1-6e^{-5})^{3/2}}{\sqrt{5}}$ & $\sqrt{\dfrac{-6+e^{5}}{-81+e^{5}}}\left(1+4e^{10}\frac{-75(-7662+639e^{5}+23e^{10})\beta+2(383586-30093e^{5}-1497e^{10}+4e^{15})s\gamma}{9(-81+e^{5})^2(-6+e^{5})^3}\right)$ \\
\hline
 $\sqrt{\frac{27}{2}}\frac{(1-5e^{-4})^{3/2}}{2}$ & $\sqrt{\dfrac{-5+e^{4}}{-53+e^{4}}}\left(1+4e^{8}\frac{-48(-419+50e^{4}+e^{8})\beta+(22435-2421e^{4}-111e^{8}+e^{12})s\gamma}{9(-53+e^{4})^2(-5+e^{4})^3}\right)$ \\
\hline
\end{tabular}
\end{table}

\begin{table}[htbp]
\centering
\caption{Critical impact parameter $u_{\rm ph}$ for different values of $\ell$, including the leading-order EFT corrections, with $M=1$.}
\label{tab:uph}
\begin{tabular}{c|c}
\hline
$\ell$ & $u_{\rm ph}$ \\
\hline
$0$ & $3\sqrt{3}-\dfrac{4s\gamma}{\sqrt{3}}$\\
\hline
$\sqrt{\frac{27}{2}}\frac{(1-101e^{-100})^{3/2}}{10}$ & $3\sqrt{\frac{3}{-301+e^{100}}}\frac{(-101+e^{100})^{3/2}}{e^{100}}+4e^{100}\sqrt{\frac{-101+e^{100}}{-903+e^{100}}}\frac{(7500\beta+(15101-e^{100})s\gamma)}{(-101+e^{100})^2}$ \\
\hline
 $\sqrt{\frac{27}{2}}\frac{(1-11e^{-10})^{3/2}}{\sqrt{10}}$ & $3\sqrt{\frac{3}{-31+e^{10}}}\frac{(-11+e^{10})^{3/2}}{e^{10}}+4e^{10}\sqrt{\frac{-11+e^{10}}{-93+e^{10}}}\frac{(75\beta+(161-e^{10})s\gamma)}{(-11+e^{10})^2}$ \\
\hline
 $\sqrt{\frac{27}{2}}\frac{(1-6e^{-5})^{3/2}}{\sqrt{5}}$ & $3\sqrt{\frac{3}{-16+e^5}}\frac{(-6+e^5)^{3/2}}{e^5}+e^5\sqrt{\frac{-6+e^5}{-48+e^5}}\frac{(75\beta+2(87-2e^5)s\gamma)}{(-6+e^5)^2}$ \\
\hline
 $\sqrt{\frac{27}{2}}\frac{(1-5e^{-4})^{3/2}}{2}$ & $3\sqrt{\frac{3}{-13+e^4}}\frac{(-5+e^4)^{3/2}}{e^4}+4e^4\sqrt{\frac{-5+e^4}{-39+e^4}}\frac{(12\beta+(29-e^4)s\gamma)}{(-5+e^4)^2}$ \\
\hline
\end{tabular}
\end{table}

The regular contribution $b_R$ and the resulting strong-deflection coefficient $\bar b$ are summarized in Table~\ref{tab:b}. For $\ell\rightarrow0$, both $b_R$ and $\bar b$ are obtained analytically, whereas for $\ell\neq0$, $b_R$ is evaluated numerically and $\bar b$ is subsequently determined from the numerical result.
\begin{table}[htbp]
\centering
\caption{Numerical values of the regular contribution $b_R$ and the strong-deflection coefficient $\bar{b}$ for different values of $\ell$, with the background contribution $b_R^{(0)}$ and the EFT corrections $b_R^{(\beta)}$, $b_R^{(\gamma)}$ proportional to $\beta$ and $s\gamma$ shown separately, with $M=1$.}
\label{tab:b}
\begin{tabular}{c|c|c|c|c}
\hline
$\ell$ & $b_R^{(0)}$ & $b_R^{(\beta)}$ & $b_R^{(\gamma)}$ & $\bar{b}$\\
\hline
$0$ & $0.949603$ & $0$ & $-1.54627$ & $-0.40023+0.138953s\gamma$\\
\hline
$\sqrt{\frac{27}{2}}\frac{(1-101e^{-100})^{3/2}}{10}$ & 0.94960 & $1.697402\times10^{-35}$ & -1.54627 &$-0.40023+1.773211\time10^{-35}\beta+0.138956s\gamma$ \\
\hline
 $\sqrt{\frac{27}{2}}\frac{(1-11e^{-10})^{3/2}}{\sqrt{10}}$ & 0.91651 & 0.63206 & -0.52919 & $-0.433889+0.623383\beta+1.16347s\gamma$\\
\hline
 $\sqrt{\frac{27}{2}}\frac{(1-6e^{-5})^{3/2}}{\sqrt{5}}$ & -0.94536 & 25.91808 & 26.949638 & $-2.464391+29.6122\beta+33.093s\gamma$\\
\hline
 $\sqrt{\frac{27}{2}}\frac{(1-5e^{-4})^{3/2}}{2}$ & -32.46132 & 58588.2 & 64087.8 & $-43.7779+79334\beta+86790s\gamma$\\
\hline
\end{tabular}
\end{table}

\begin{figure}[t]
    \centering

    \begin{subfigure}{0.32\textwidth}
        \centering
        \includegraphics[width=\textwidth]{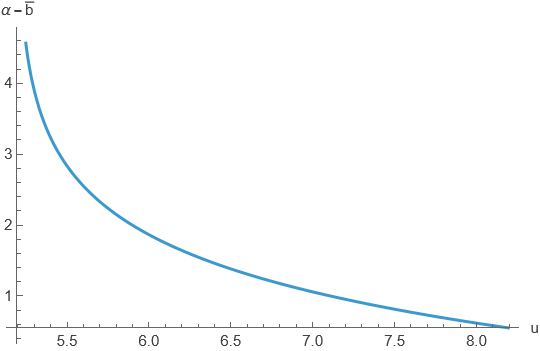}
        \caption{$Schwarzschild$}
    \end{subfigure}
    \hfill
    \begin{subfigure}{0.32\textwidth}
        \centering
        \includegraphics[width=\textwidth]{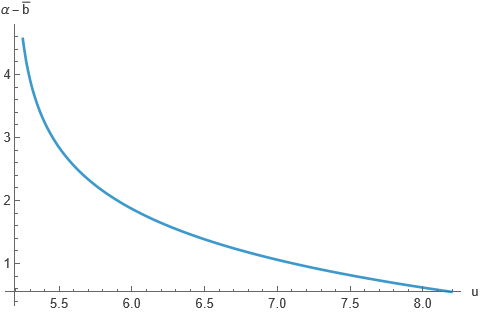}
        \caption{$\ell=\sqrt{\frac{27}{2}}\frac{(1-101e^{-100})^{3/2}}{10}$}
    \end{subfigure}
    \hfill
    \begin{subfigure}{0.32\textwidth}
        \centering
        \includegraphics[width=\textwidth]{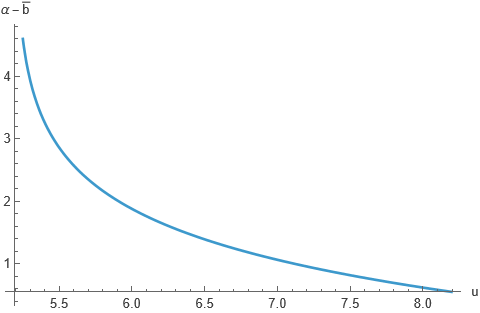}
        \caption{$\ell=\sqrt{\frac{27}{2}}\frac{(1-11e^{-10})^{3/2}}{\sqrt{10}}$}
    \end{subfigure}
    \hfill
    \begin{subfigure}{0.32\textwidth}
        \centering
        \includegraphics[width=\textwidth]{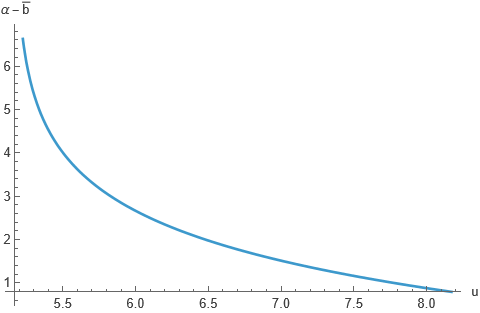}
        \caption{$\ell=\sqrt{\frac{27}{2}}\frac{(1-6e^{-5})^{3/2}}{\sqrt{5}}$}
    \end{subfigure}
    \hfill
    \begin{subfigure}{0.32\textwidth}
        \centering
        \includegraphics[width=\textwidth]{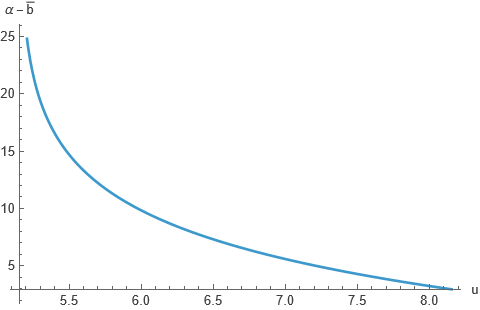}
        \caption{$\ell=\sqrt{\frac{27}{2}}\frac{(1-5e^{-4})^{3/2}}{2}$}
    \end{subfigure}

    \caption{Deflection angle $\alpha(u)$ for the higher-curvature-corrected Dymnikova black hole and higher-curvature coupling constants $\beta=\gamma=10^{-5}$.}
    \label{fig:RB0_deflection_angle}
\end{figure}

\begin{figure}[t]
    \centering

    \begin{subfigure}{0.32\textwidth}
        \centering
        \includegraphics[width=\textwidth]{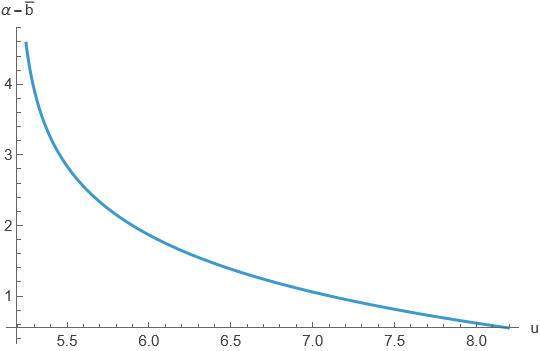}
        \caption{$Schwarzschild$}
    \end{subfigure}
    \hfill
\begin{subfigure}{0.32\textwidth}
        \centering
        \includegraphics[width=\textwidth]{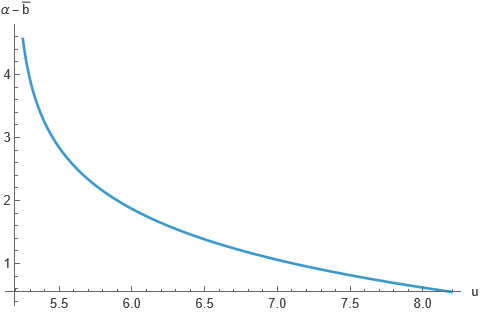}
        \caption{$\ell=\sqrt{\frac{27}{2}}\frac{(1-101e^{-100})^{3/2}}{10}$}
    \end{subfigure}
    \hfill
    \begin{subfigure}{0.32\textwidth}
        \centering
        \includegraphics[width=\textwidth]{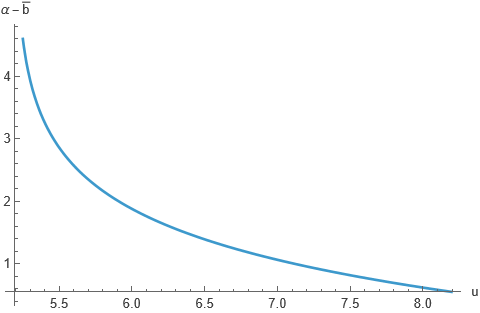}
        \caption{$\ell=\sqrt{\frac{27}{2}}\frac{(1-11e^{-10})^{3/2}}{\sqrt{10}}$}
    \end{subfigure}
    \hfill
    \begin{subfigure}{0.32\textwidth}
        \centering
        \includegraphics[width=\textwidth]{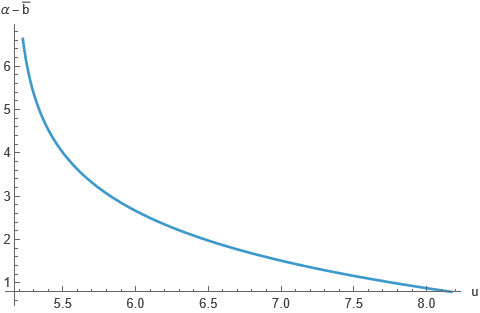}
        \caption{$\ell=\sqrt{\frac{27}{2}}\frac{(1-6e^{-5})^{3/2}}{\sqrt{5}}$}
    \end{subfigure}
    \hfill
    \begin{subfigure}{0.32\textwidth}
        \centering
        \includegraphics[width=\textwidth]{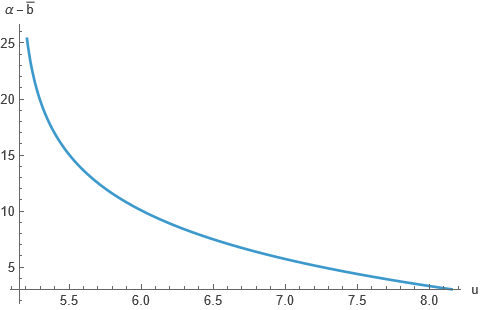}
        \caption{$\ell=\sqrt{\frac{27}{2}}\frac{(1-5e^{-4})^{3/2}}{2}$}
    \end{subfigure}

    \caption{Deflection angle $\alpha(u)$ for the Hayward black hole without EFT corrections.}
    \label{fig:RB00}
\end{figure}

For each of the five representative values of $\ell$ considered above, we compare the deflection angles with and without the EFT corrections. For $\ell\rightarrow0$, corresponding to the Schwarzschild limit, the full strong-deflection angle can be obtained analytically and is used as a reference for the comparison. The corresponding results are shown in Fig.~\ref{fig:RB0_deflection_angle}. For the remaining four nonzero values of $\ell$, we focus on the logarithmically divergent part of the deflection angle, since the perturbative EFT corrections can become more pronounced in the strong-deflection regime. The corresponding results are shown in Fig.~\ref{fig:RB00}. The EFT coupling constants are fixed to $\beta=\gamma=10^{-5}$ throughout the numerical analysis.

\begin{figure}[t]
    \centering

  \begin{subfigure}{0.32\textwidth}
        \centering
        \includegraphics[width=\textwidth]{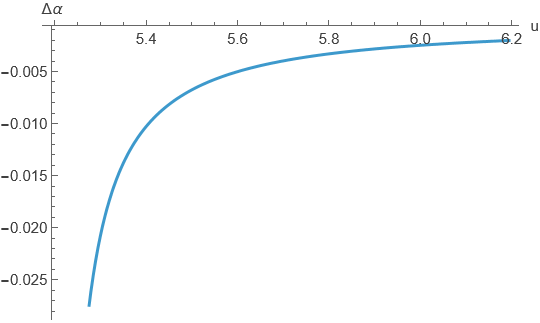}
        \caption{$Schwarzschild$}
    \end{subfigure}
    \hfill  
\begin{subfigure}{0.32\textwidth}
        \centering
        \includegraphics[width=\textwidth]{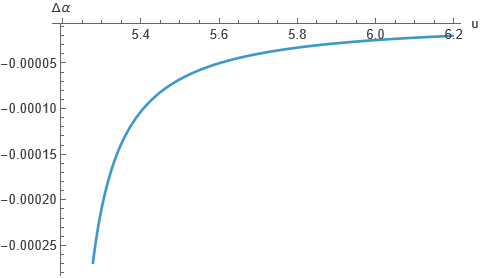}
        \caption{$\ell=\sqrt{\frac{27}{2}}\frac{(1-101e^{-100})^{3/2}}{10}$}
    \end{subfigure}
    \hfill
  \begin{subfigure}{0.32\textwidth}
        \centering
        \includegraphics[width=\textwidth]{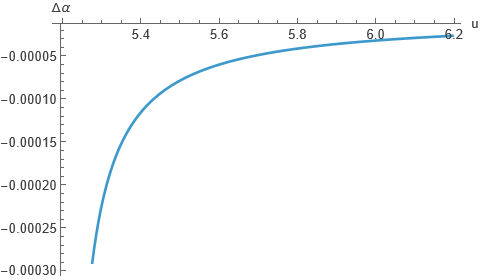}
        \caption{$\ell=\sqrt{\frac{27}{2}}\frac{(1-11e^{-10})^{3/2}}{\sqrt{10}}$}
    \end{subfigure}
    \hfill
    \begin{subfigure}{0.32\textwidth}
        \centering
        \includegraphics[width=\textwidth]{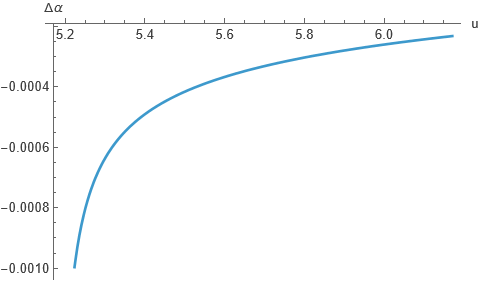}
        \caption{$\ell=\sqrt{\frac{27}{2}}\frac{(1-6e^{-5})^{3/2}}{\sqrt{5}}$}
    \end{subfigure}
    \hfill
    \begin{subfigure}{0.32\textwidth}
        \centering
        \includegraphics[width=\textwidth]{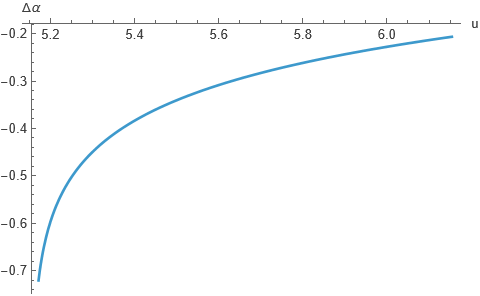}
        \caption{$\ell=\sqrt{\frac{27}{2}}\frac{(1-5e^{-4})^{3/2}}{2}$}
    \end{subfigure}

    \caption{Difference in the deflection angle $\Delta\alpha$ between the higher-curvature-corrected Dymnikova black hole and the Dymnikova black hole without EFT corrections for higher-curvature coupling constants $\beta=\gamma=10^{-5}$.}
    \label{fig:RB_deflection_delta}
\end{figure}

For each value of $\ell$, we also consider the difference between the deflection angles with and without the EFT corrections,
\begin{align}
\Delta\alpha = \alpha_{\rm EFT}-\alpha_{\rm 0},
\end{align}
where $\alpha_{\rm EFT}$ and $\alpha_{\rm 0}$ denote the deflection angles with and without the EFT corrections, respectively. The difference $\Delta\alpha$ isolates the effect of the EFT corrections at each fixed value of $\ell$. The corresponding results are shown in Fig.~\ref{fig:RB_deflection_delta}, allowing us to assess the contribution of the EFT corrections to the strong-deflection behavior.


The strong-deflection limit corresponds to the case in which the closest approach $r_0$ approaches the radius of the photon sphere $r_{\rm ph}$. In this limit, the deflection angle exhibits a logarithmic divergence, which can be systematically analyzed using the formalism developed in the previous subsection. For the Dymnikova black hole considered in this work, even when the EFT corrections are parametrically small, their contributions to the logarithmically divergent term can become appreciable in the strong-deflection regime. This enhancement provides a potentially sensitive probe of curvature-dependent modifications of photon propagation in the near-photon-sphere region.

For the Schwarzschild limit, $\ell\rightarrow0$, the strong-deflection quantities can be evaluated analytically. The corresponding analytic results for the photon-sphere radius, critical impact parameter, and strong-deflection coefficients are summarized in Appendix~\ref{app:schwarzschild}. For nonzero values of $\ell$, these quantities are instead evaluated numerically.

One of the main goals of this work is to identify observational signatures that can distinguish the Dymnikova black hole from the Schwarzschild black hole through gravitational lensing. In particular, the Dymnikova spacetime approaches the Schwarzschild spacetime in the limit $\ell\rightarrow0$, while nonzero values of $\ell$ modify the geometry in the near-photon-sphere region. These geometric modifications, together with the EFT corrections, affect the photon-sphere radius, the critical impact parameter, and the coefficients governing the logarithmically divergent part of the deflection angle. Although the EFT corrections may be parametrically small, their contributions to the strong-deflection coefficients can become appreciable in the strong-deflection regime. Moreover, the deviations of the lensing observables from the corresponding Schwarzschild values become more pronounced as $\ell$ increases, both with and without the EFT corrections. The resulting modifications of the lensing observables may therefore provide characteristic signatures that distinguish the Dymnikova and Schwarzschild black holes and probe the effects of curvature-dependent corrections to photon propagation.

In this work, we plot the results by taking the EFT coupling constants to be $\beta=\gamma=10^{-5}$. The difference between the two modes enters only through the sign of the corresponding EFT coupling terms. Therefore, the numerical results are expected to be qualitatively similar for both polarization modes, with the EFT-induced deviations appearing with opposite signs. In this sense, the qualitative behavior obtained in the numerical analysis does not depend on the choice of polarization mode.

\section{Time Delay in EFT-Corrected Hayward black hole}
\label{sec:time delay}

In this section, we investigate the time delay of photons in the EFT-corrected Dymnikova black hole spacetime. As in the gravitational lensing analysis presented in the previous section, we use the time delay as an observable to explore potential differences between the Dymnikova black hole and the Schwarzschild black hole. In particular, we derive the photon time delay in the strong-deflection limit and examine how the EFT-induced modifications to photon propagation affect this observable. This analysis allows us to assess whether the time-delay signal can provide an additional observational signature for distinguishing the Dymnikova black hole from the Schwarzschild black hole.

\subsection{Travel Time Integral}

The coordinate time along a photon trajectory is obtained by eliminating the affine parameter,
\begin{equation}
\frac{dt}{dr} = \frac{\dot{t}}{\dot{r}},
\end{equation}
where $\dot{t}$ and $\dot{r}$ are determined from the conserved energy and angular momentum. Using these conserved quantities and the null condition, we obtain
\begin{align}
\frac{dt}{dr}
=&\dfrac{\sqrt{BA_0}}{\sqrt{A}}
\dfrac{1}{\sqrt{A_0-A\frac{C_0}{C}}}.
\end{align}
The photon travel time follows by integrating this expression along the trajectory. As in the strong-deflection analysis of the deflection angle, the integral can be separated into a contribution containing the singular behavior near the photon sphere and a regular contribution.
The time difference between two photon trajectories with closest approach radii $r_{0,1}$ and $r_{0,2}$ is written as
\begin{align}
T_1-T_2
=&\tilde{T}(r_{0,1})-\tilde{T}(r_{0,2})
+2\int_{r_{0,1}}^{r_{0,2}}
\frac{\sqrt{B(r)}}{\sqrt{A(r)}},dr,
\end{align}
where
\begin{align}
\tilde{T}(r_0)
=&\int_0^1
\tilde{R}
\frac{1}{\sqrt{A_0-A(r(z))\frac{C_0}{C(r(z))}}},dz,\\
\tilde{R}(z,r_0)
=&2\frac{r_0}{(1-z)^2}
\frac{\sqrt{B(r(z))A_0}}{\sqrt{A(r(z))}}
\left(
1-
\frac{
\sqrt{A_0-A(r(z))\frac{C_0}{C(r(z))}}
}{
\sqrt{A_0}
}
\right).
\end{align}
Here, $z$ is defined by Eq.~(\ref{zcoor}), with $z=0$ corresponding to the closest approach. The logarithmic behavior near the photon sphere is contained in the factor $A_0-A(r(z))C_0/C(r(z))$.
Expanding the integrand around the photon sphere yields the strong-deflection form
\begin{align}
\tilde{T}(u)
=-\tilde{a}\log\left(\frac{u}{u_{\rm ph}}-1\right)+\tilde{b}+\mathcal{O}\left((u-u_{\rm ph})\log(u-u_{\rm ph})\right),
\end{align}
where $u_{\rm ph}$ is the critical impact parameter. The coefficients $\tilde{a}$ and $\tilde{b}$ characterize the leading behavior of the photon travel time in the strong-deflection limit, with $\tilde{a}$ determining the coefficient of the leading logarithmic divergence and given by
\begin{align}
\tilde{a}
=\dfrac{\tilde{R}(0,r_{\rm ph})}{2\sqrt{q_{\rm ph}}}.
\end{align}
The coefficient $\tilde{b}$ corresponds to the constant term with respect to $u$ in the strong-deflection expansion.

\subsection{Time Delay in the Dymnikova Spacetime}

We apply the above formalism to the five representative values of $\ell$ considered in the previous section. The corresponding photon-sphere properties and strong-deflection coefficients are used to evaluate the photon travel time and time delay.
For the Schwarzschild limit, the strong-deflection time delay can be obtained analytically. For the remaining three representative values of $\ell$, the travel-time integral is evaluated numerically, with particular attention to its behavior near the photon sphere.
The resulting strong-deflection observables are summarized in Table~\ref{tab:tilde_a} and \ref{tab:tilde_b}. The Schwarzschild case provides the reference result, while the other cases characterize the effect of the Dymnikova parameter on the photon propagation and the associated time delay.
\begin{align}
\tilde{b}=&-\pi+12-6\sqrt{3}-12\sqrt{3}\arctan(\sqrt{3})+3\sqrt{3}\log6+\sqrt{3}\log46656+\log(58406416-33720960\sqrt{3})\notag\\
&+\frac{8}{3}(-3+\sqrt{3})s\gamma,\\
\tilde{T}(u)=&-3\sqrt{3}\log\left(\frac{u}{3\sqrt{3}-\frac{4s\gamma}{\sqrt{3}}}-1\right)+\tilde{b}+\mathcal{O}(u-u_{\rm ph}).
\end{align}

For each of the five representative values of $\ell$ considered above, we investigate the photon travel time in the strong-deflection regime and compare the results with and without the EFT corrections. In the Schwarzschild limit, $\ell\rightarrow0$, the strong-deflection expansion of the photon travel time can be obtained analytically. For the remaining four nonzero values of $\ell$, we numerically evaluate the photon travel time, focusing in particular on its logarithmically divergent behavior near the photon sphere, where the effects of the perturbative EFT corrections can become more pronounced.

\begin{figure}[t]
    \centering

    \begin{subfigure}{0.32\textwidth}
        \centering
        \includegraphics[width=\textwidth]{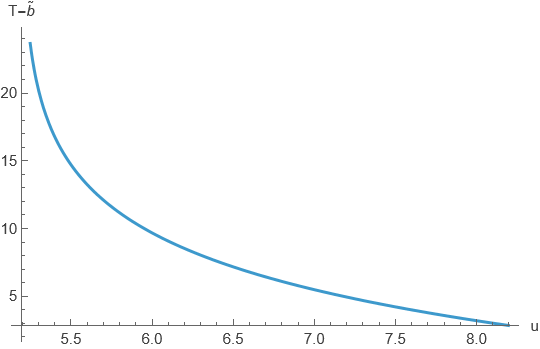}
        \caption{Schwarzschild case}
    \end{subfigure}
    \hfill
    \begin{subfigure}{0.32\textwidth}
        \centering
        \includegraphics[width=\textwidth]{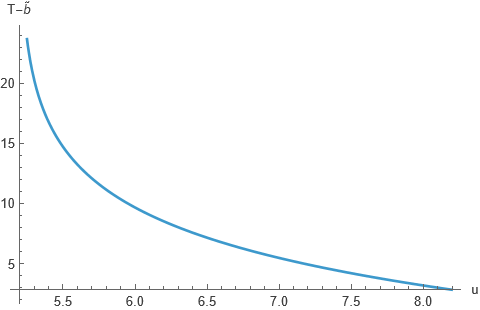}
        \caption{$\ell=\sqrt{\frac{27}{2}}\frac{(1-101e^{-100})^{3/2}}{10}$}
    \end{subfigure}
    \hfill
    \begin{subfigure}{0.32\textwidth}
        \centering
        \includegraphics[width=\textwidth]{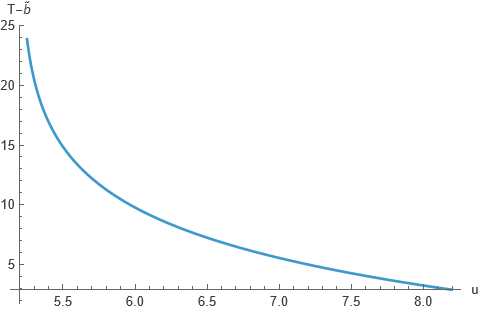}
        \caption{$\ell=\sqrt{\frac{27}{2}}\frac{(1-11e^{-10})^{3/2}}{\sqrt{10}}$}
    \end{subfigure}
    \hfill
    \begin{subfigure}{0.32\textwidth}
        \centering
        \includegraphics[width=\textwidth]{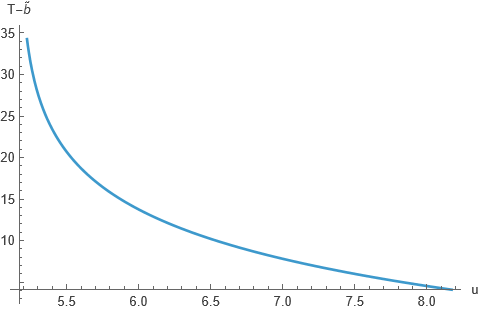}
        \caption{$\ell=\sqrt{\frac{27}{2}}\frac{(1-6e^{-5})^{3/2}}{\sqrt{5}}$}
    \end{subfigure}
    \hfill
    \begin{subfigure}{0.32\textwidth}
        \centering
        \includegraphics[width=\textwidth]{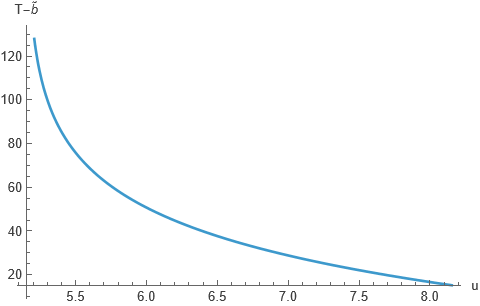}
        \caption{$\ell=\sqrt{\frac{27}{2}}\frac{(1-5e^{-4})^{3/2}}{2}$}
    \end{subfigure}

    \caption{Photon travel time $T(u)$ for the higher-curvature-corrected Dymnikova black hole and higher-curvature coupling constants $\beta=\gamma=10^{-5}$.}
    \label{fig:RB0_time_delay}
\end{figure}

\begin{figure}[t]
    \centering

    \begin{subfigure}{0.32\textwidth}
        \centering
        \includegraphics[width=\textwidth]{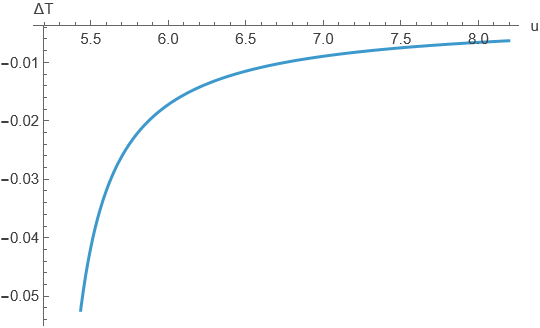}
        \caption{Schwarzschild case}
    \end{subfigure}
    \hfill
    \begin{subfigure}{0.32\textwidth}
        \centering
        \includegraphics[width=\textwidth]{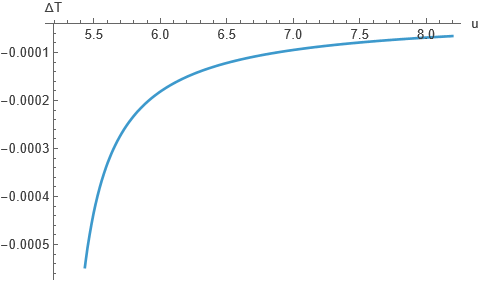}
        \caption{$\ell=\sqrt{\frac{27}{2}}\frac{(1-101e^{-100})^{3/2}}{10}$}
    \end{subfigure}
    \hfill
   \begin{subfigure}{0.32\textwidth}
        \centering
        \includegraphics[width=\textwidth]{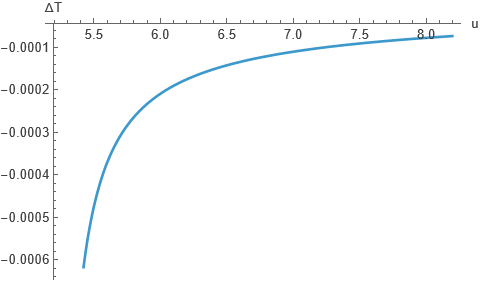}
        \caption{$\ell=\sqrt{\frac{27}{2}}\frac{(1-11e^{-10})^{3/2}}{\sqrt{10}}$}
    \end{subfigure}
    \hfill
    \begin{subfigure}{0.32\textwidth}
        \centering
        \includegraphics[width=\textwidth]{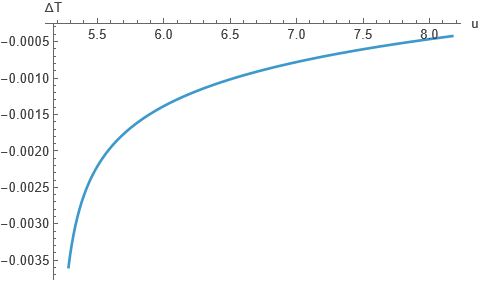}
        \caption{$\ell=\sqrt{\frac{27}{2}}\frac{(1-6e^{-5})^{3/2}}{\sqrt{5}}$}
    \end{subfigure}
    \hfill
    \begin{subfigure}{0.32\textwidth}
        \centering
        \includegraphics[width=\textwidth]{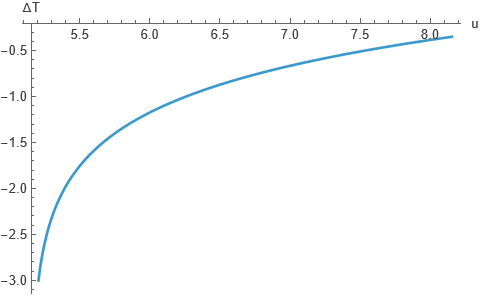}
        \caption{$\ell=\sqrt{\frac{27}{2}}\frac{(1-5e^{-4})^{3/2}}{2}$}
    \end{subfigure}

    \caption{Difference in the photon travel time, $\Delta T(u)$, between the higher-curvature-corrected Dymnikova black hole and the Dymnikova black hole without EFT corrections, for higher-curvature coupling constants $\beta=\gamma=10^{-5}$.}
    \label{fig:RB_time_delay_delta}
\end{figure}

The photon travel times including the EFT corrections are shown in Fig.~\ref{fig:RB0_time_delay}, while the corresponding differences between the EFT-corrected and uncorrected cases are presented in Fig.~\ref{fig:RB_time_delay_delta}. Throughout the numerical analysis, the EFT coupling constants are fixed to $\beta=\gamma=10^{-5}$.
For each value of $\ell$, we also consider the difference between the photon travel times with and without the EFT corrections,
\begin{align}
\Delta T = T_{\rm EFT}-T_0,
\end{align}
where $T_{\rm EFT}$ and $T_0$ denote the photon travel times with and without the EFT corrections, respectively. This quantity isolates the effect of the EFT corrections for a fixed value of $\ell$. The resulting differences are shown in Fig.~\ref{fig}, providing a direct measure of the EFT contribution to the strong-deflection behavior.

\begin{table}[htbp]
\centering
\caption{Strong-deflection coefficient $\tilde{a}$ for different values of $\ell$, including the leading-order EFT corrections, with $M=1$.}
\label{tab:tilde_a}
\begin{tabular}{c|c}
\hline
$\ell$ & $\tilde{a}$ \\
\hline
$0$ & $3\sqrt{3}$\\
\hline
$\sqrt{\frac{27}{2}}\frac{(1-101e^{-100})^{3/2}}{10}$ & $\sqrt{\frac{3(-301+e^{100})}{-30101+e^{100}}}\left(3\frac{(-101+e^{100})^2}{e^{100}(-301+e^{100})}-\frac{10000e^{100}((4530297+14403e^{100})\beta+(9060186+28414e^{100})s\gamma)}{(-30101+e^{100})^2(-101+e^{100})}\right)$ \\
\hline
$\sqrt{\frac{27}{2}}\frac{(1-11e^{-10})^{3/2}}{\sqrt{10}}$ & $\sqrt{\frac{3(-31+e^{10})}{-311+e^{10}}}\left(3\frac{(-11+e^{10})^2}{e^{10}(-31+e^{10})}-\frac{100e^{10}((4827+93e^{10})\beta+2(4803+77e^{10})s\gamma)}{(-311+e^{10})^2(-11+e^{10})}\right)$ \\
\hline
$\sqrt{\frac{27}{2}}\frac{(1-6e^{-5})^{3/2}}{\sqrt{5}}$ & $\sqrt{\frac{3(-16+e^{5})}{-81+e^{5}}}\left(3\frac{(-6+e^{5})^2}{e^{5}(-16+e^{5})}-\frac{25e^{5}(3(433+7e^{5})\beta+2(1271+9e^{5})s\gamma)}{2(-81+e^{5})^2(-6+e^{5})}\right)$ \\
\hline
$\sqrt{\frac{27}{2}}\frac{(1-5e^{-4})^{3/2}}{2}$ & $\sqrt{\frac{3(-13+e^{4})}{-53+e^{4}}}\left(3\frac{(-5+e^{4})^2}{e^{4}(-13+e^{5})}-\frac{16e^{4}(3(115+e^{4})\beta+2(333-e^{4})s\gamma)}{(-53+e^{4})^2(-5+e^{4})}\right)$ \\
\hline
\end{tabular}
\end{table}

\begin{figure}[t]
    \centering

     \begin{subfigure}{0.32\textwidth}
        \centering
        \includegraphics[width=\textwidth]{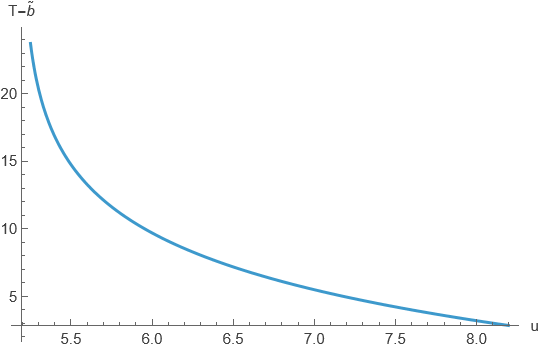}
        \caption{Schwarzschild case}
    \end{subfigure}
    \hfill
    \begin{subfigure}{0.32\textwidth}
        \centering
        \includegraphics[width=\textwidth]{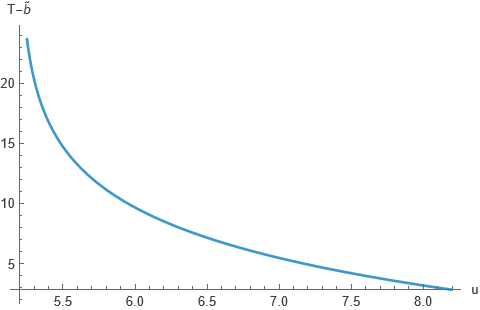}
        \caption{$\ell=\sqrt{\frac{27}{2}}\frac{(1-101e^{-100})^{3/2}}{10}$}
    \end{subfigure}
    \hfill
    \begin{subfigure}{0.32\textwidth}
        \centering
        \includegraphics[width=\textwidth]{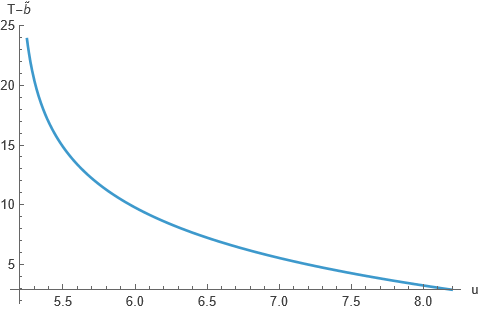}
        \caption{$\ell=\sqrt{\frac{27}{2}}\frac{(1-11e^{-10})^{3/2}}{\sqrt{10}}$}
    \end{subfigure}
    \hfill
    \begin{subfigure}{0.32\textwidth}
        \centering
        \includegraphics[width=\textwidth]{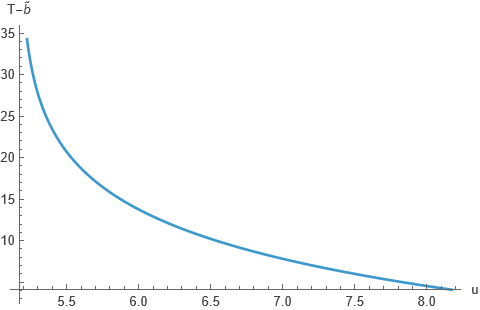}
        \caption{$\ell=\sqrt{\frac{27}{2}}\frac{(1-6e^{-5})^{3/2}}{\sqrt{5}}$}
    \end{subfigure}
    \hfill
    \begin{subfigure}{0.32\textwidth}
        \centering
        \includegraphics[width=\textwidth]{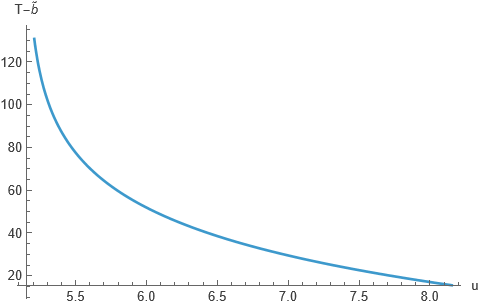}
        \caption{$\ell=\sqrt{\frac{27}{2}}\frac{(1-5e^{-4})^{3/2}}{2}$}
    \end{subfigure}

    \caption{Photon travel time $T(u)$ for the Dymnikova black hole without EFT corrections.}
    \label{fig:RB00_time_delay}
\end{figure}
\subsection{Time Delay Between Relativistic Images}

In the strong-deflection regime, relativistic images correspond to photon trajectories that wind multiple times around the photon sphere of the Dymnikova black hole. For trajectories with winding number $n$, the corresponding impact parameters satisfy
\begin{align}
\frac{u}{u_{\rm ph}} - 1 \simeq \exp\left(\frac{\bar{b} - 2\pi n}{\bar{a}}\right).
\end{align}

Substituting this relation into the time expression, we obtain
\begin{align}
T_n = \tilde{a} \frac{2\pi n - \bar{b}}{\bar{a}} + \tilde{b}.
\end{align}
Here, $T_n$ denotes the photon travel time for a trajectory that winds $n$ times around the photon sphere.

The time delay between two relativistic images labeled by $n$ and $m$ is therefore given by
\begin{align}
\Delta T_{n,m}\equiv& \ T_n-T_m\notag\\
=&\ 2\pi(n-m)\left|\frac{\tilde{a}}{\bar{a}}\right|+2\sqrt{\dfrac{B_{\rm ph}}{A_{\rm ph}}}\sqrt{\dfrac{u_{\rm ph}}{c}}e^{\frac{\bar{b}}{2\bar{a}}}\left(e^{-\frac{\pi n}{|\bar{a}|}}-e^{\frac{\pi m}{|\bar{a}|}}\right)\notag\\
\equiv&\ \tilde{T}_{1}(n-m)+\tilde{T}_{2}\left(e^{-\frac{\pi n}{|\bar{a}|}}-e^{\frac{\pi m}{|\bar{a}|}}\right),
\end{align}
where $\tilde{T}_{1}$ and $\tilde{T}_{2}$ are defined by
\begin{align}
\tilde{T}_{1}=&2\pi\left|\frac{\tilde{a}}{\bar{a}}\right|,\\
\tilde{T}_{2}=&2\sqrt{\dfrac{B_{\rm ph}}{A_{\rm ph}}}\sqrt{\dfrac{u_{\rm ph}}{c}}e^{\frac{\bar{b}}{2\bar{a}}},
\end{align}
where $c$ is the quantity efined in Eq.~(\ref{c_def}).

We now turn to the numerical evaluation of the time-delay observable for the EFT-corrected Dymnikova black hole. Throughout this analysis, we fix the mass parameter to $M=1$. For each of the PPL and PPM polarization modes, the quantities $\tilde{T}_1$ and $\tilde{T}_2$ are calculated numerically for several representative values of the Dymnikova parameter $\ell$. The resulting values are listed in Tables~\ref{tab:T1} and \ref{tab:T2}.

For the numerical analysis presented in Figs.~\ref{fig:RB0_time_delay} and \ref{fig:RB_time_delay_delta}, we plot the results by taking the EFT coupling constants to be $\beta=\gamma=10^{-5}$. The difference between the two modes enters primarily through the opposite signs of the corresponding EFT correction terms. Therefore, we take $\beta=\gamma=10^{-5}$ as representative values of the EFT coupling constants in the numerical analysis. Since the polarization dependence is encoded only through the sign of the EFT contributions, the two modes are expected to exhibit qualitatively similar behavior, with the corresponding EFT-induced deviations appearing with opposite signs.

We also find that the deviation of the time-delay observable from the Schwarzschild result becomes more pronounced as the Dymnikova parameter $\ell$ increases, both with and without the EFT corrections. In addition, the magnitude of the EFT corrections themselves increases with $\ell$. Thus, for larger $\ell$, the time-delay observable exhibits a larger departure from the Schwarzschild prediction, while the difference between the EFT-corrected and uncorrected results also becomes more significant. This indicates that the effects of the Dymnikova geometry and the EFT corrections can become increasingly relevant for distinguishing the Dymnikova and Schwarzschild spacetimes.

It is also useful to consider the behavior of the photon travel time itself in the vicinity of the critical impact parameter. As the impact parameter approaches its critical value, the travel time grows logarithmically, reflecting the fact that photons spend an increasingly long time near the unstable photon orbit while making repeated windings around the photon sphere. Although this travel time is not itself a directly measurable quantity, its divergent behavior determines the contribution to the time difference between photons propagating along trajectories with different winding numbers. The resulting time delay consequently carries information about the winding structure of the photon trajectories.

The EFT corrections modify this strong-deflection behavior and hence affect the winding-dependent time delay. Their impact becomes more evident for larger values of $\ell$, leading to a larger difference from the Schwarzschild result. These results demonstrate that the photon time delay can serve as a complementary probe to the deflection angle: while the two observables contain different information about photon propagation, their combined behavior can provide a more complete assessment of the differences between the Dymnikova and Schwarzschild black holes and of the effects induced by the EFT corrections.

\begin{table}[htbp]
\centering
\caption{Values of $\tilde{T}_1$ for representative values of the parameter $\ell$.}
\label{tab:T1}
\begin{tabular}{c|c}
\hline
$\ell$ & $\tilde{T}_1$ \\
\hline
$0$ & $6\sqrt{3}\pi-\dfrac{8s\gamma\pi}{\sqrt{3}}$ \\
\hline
$\sqrt{\frac{27}{2}}\frac{(1-101e^{-100})^{3/2}}{10}$ & $6\sqrt{\frac{3(-101+e^{100})}{-301+e^{100}}}\frac{(-101+e^{100})\pi}{e^{100}}+8e^{100}\sqrt{\frac{-301+e^{100}}{3(-101+e^{10})}}\pi\frac{7500\beta+(15101-e^{100})s\gamma}{(301-e^{100})(101-e^{100})}$ \\
\hline
$\sqrt{\frac{27}{2}}\frac{(1-11e^{-10})^{3/2}}{\sqrt{10}}$ & $6\sqrt{\frac{3(-11+e^{10})}{-31+e^{10}}}\frac{(-11+e^{10})\pi}{e^{10}}+8e^{10}\sqrt{\frac{-31+e^{10}}{3(-11+e^{10})}}\pi\frac{75\beta+(161-e^{10})s\gamma}{(31-e^{10})(11-e^{10})}$ \\
\hline
$\sqrt{\frac{27}{2}}\frac{(1-6e^{-5})^{3/2}}{\sqrt{5}}$ &  $6\sqrt{\frac{3(-6+e^5)}{-16+e^5}}\frac{(-6+e^5)\pi}{e^5}+2e^5\sqrt{\frac{-6+e^5}{3(-6+e^4)}}\pi\frac{75\beta+(87-2e^5)s\gamma}{(-6+e^5)^2}$  \\
\hline
$\sqrt{\frac{27}{2}}\frac{(1-5e^{-4})^{3/2}}{2}$ & $6\sqrt{\frac{3(-5+e^4)}{-13+e^4}}\frac{(-5+e^4)\pi}{e^4}+8e^4\sqrt{\frac{-13+e^4}{3(-5+e^4)}}\pi\frac{12\beta+(29-e^4)s\gamma}{(13-e^4)(5-e^4)}$\\
\hline
\end{tabular}
\end{table}
\begin{table}[htbp]
\centering
\caption{Values of $\tilde{T}_2$ for representative values of the parameter $\ell$.}
\label{tab:T2}
\begin{tabular}{c|c}
\hline
$\ell$ & $\tilde{T}_2$ \\
\hline
$0$ & \makecell[l]{
$6\sqrt{6}e^{\frac{1}{2}\left(-\pi-4\arctan\left(\frac{1}{\sqrt{3}}\right)+\log216\right)}+\frac{4}{5}\sqrt{\frac{2}{3}}e^{\frac{1}{2}\left(-\pi-4\arctan\left(\frac{1}{\sqrt{3}}\right)+\log216\right)}s\gamma$\\
$\times\left(-26+8\sqrt{3}+5\pi+20\arctan\left(\frac{1}{\sqrt{3}}\right)+\log1024-20\log(1+\sqrt{3})\right)$}\\
\hline
$\sqrt{\frac{27}{2}}\frac{(1-101e^{-100})^{3/2}}{10}$ & $12.0314+7.84104\times10^{-35}\beta+1.90601s\gamma$ \\
\hline
$\sqrt{\frac{27}{2}}\frac{(1-11e^{-10})^{3/2}}{\sqrt{10}}$ & $11.9292+1.64615\beta+4.54113s\gamma$ \\
\hline
$\sqrt{\frac{27}{2}}\frac{(1-6e^{-5})^{3/2}}{\sqrt{5}}$ & $9.10767+23.4952\beta+25.2084s\gamma$ \\
\hline
$\sqrt{\frac{27}{2}}\frac{(1-5e^{-4})^{3/2}}{2}$ & $1.59668+2869.91\beta+3139.19s\gamma$
\\
\hline
\end{tabular}
\end{table}

\section{Conclusion and Discussion}
\label{sec:conclusion}

In this work, we have investigated whether a Dymnikova regular black hole can be distinguished from a Schwarzschild black hole through strong gravitational lensing when effective field theory (EFT) corrections to photon propagation are taken into account. Starting from an effective action containing curvature-photon couplings, we derived the modified photon propagation law and constructed the corresponding polarization-dependent effective optical metrics. This provides a framework for studying strong gravitational lensing and photon time delay in the Dymnikova spacetime while consistently incorporating curvature-dependent corrections to photon propagation.

We first analyzed the strong-deflection behavior of photon trajectories in the Dymnikova spacetime and obtained the photon-sphere radius, critical impact parameter, and strong-deflection coefficients. We then incorporated the EFT corrections and examined their effects on these quantities for the different photon polarization modes. Although the EFT corrections are perturbatively small, their effects can be enhanced in the strong-deflection regime, where the deflection angle exhibits a logarithmic divergence as the impact parameter approaches its critical value. We found that the EFT corrections modify the strong-deflection observables in a polarization-dependent manner. By considering representative values of the Dymnikova parameter $\ell$ and comparing the resulting observables with those of the Schwarzschild black hole, we investigated the extent to which the combined effects of the Dymnikova geometry and EFT-corrected photon propagation can lead to distinguishable lensing signatures.

An important aspect of the present analysis is the non-Ricci-flat nature of the Dymnikova spacetime. In contrast to the Schwarzschild spacetime, where the Ricci tensor vanishes in vacuum, the Ricci curvature of the Dymnikova geometry is generally nonzero. Consequently, curvature-photon interactions involving the Ricci tensor can contribute to photon propagation in the Dymnikova spacetime, in addition to the Weyl-tensor contribution. These curvature-dependent corrections therefore probe the specific curvature structure of the regular black hole and do not simply result in a universal shift of the lensing observables. The polarization dependence of the effective optical metric further leads to different photon trajectories and strong-deflection observables for the two physical polarization modes. This demonstrates that strong gravitational lensing can probe not only deviations of the background geometry from Schwarzschild but also the curvature-dependent modification of photon propagation.

We have also examined the photon time delay as a complementary observable. In the strong-deflection regime, the photon travel time exhibits a logarithmic divergence as the impact parameter approaches its critical value, reflecting the increasingly long propagation of photons in the vicinity of the unstable photon orbit. Although the absolute photon travel time is not directly observable, the time delay between photons following trajectories with different winding numbers can be constructed from differences in their travel times. The logarithmic behavior of the travel time therefore provides the underlying contribution to the winding-dependent time-delay signal. We find that the EFT corrections also modify the strong-deflection behavior of the photon travel time, with the corresponding effects depending on the photon polarization and the Dymnikova parameter. The time delay thus provides a complementary probe of the differences between the Dymnikova and Schwarzschild spacetimes.

Our results further show that the deviations of the strong-deflection observables from their Schwarzschild values become more pronounced as the Dymnikova parameter $\ell$ increases, both with and without the EFT corrections. In particular, the difference between the Dymnikova and Schwarzschild results becomes increasingly significant for larger $\ell$, indicating that the deviation from the Schwarzschild spacetime is enhanced by the Dymnikova geometry itself. In addition, the magnitude of the EFT corrections also increases with $\ell$, leading to a more pronounced difference between the EFT-corrected and uncorrected results. This behavior is found in both the strong-deflection angle and the photon time delay, although the magnitude and polarization dependence of the corrections differ between the observables. Thus, the Dymnikova parameter $\ell$ plays an important role in determining both the deviation from the Schwarzschild predictions and the sensitivity of strong gravitational lensing to curvature-dependent modifications of photon propagation.

The dependence on $\ell$ is particularly relevant when assessing the observational distinguishability of the Dymnikova black hole. As $\ell$ increases, the deviation from the Schwarzschild geometry becomes more pronounced, while the curvature-dependent EFT corrections to photon propagation also become increasingly relevant. The combined effects therefore lead to larger deviations in the strong-deflection observables from the Schwarzschild predictions. This suggests that strong gravitational lensing and photon time delay can provide sensitive probes of the Dymnikova parameter and the EFT corrections, particularly for configurations with relatively large $\ell$.

Several directions remain for future investigation. First, it would be interesting to extend the present analysis to other regular black hole spacetimes, such as the Bardeen~\cite{1968qtr..conf...87B} and Ay'on-Beato--Garc'ia black holes~\cite{Ayon-Beato:1998hmi,Ayon-Beato:1999kuh}, and to investigate how their different curvature structures affect the EFT corrections to photon propagation and the resulting strong-deflection observables. Such a comparison would help clarify which features of the lensing and time-delay signals are specific to the Dymnikova geometry and which are common to regular black holes more generally. It would also be valuable to incorporate finite-distance effects and higher-order terms in the strong-deflection expansion, as well as to systematically explore the allowed parameter space of the Dymnikova and EFT couplings. Finally, confronting the predicted polarization-dependent lensing and time-delay signatures with current and future high-precision observations could provide a more direct assessment of the observational prospects for distinguishing regular black holes from the Schwarzschild black hole and for probing curvature-dependent corrections to photon propagation.

\appendix

\section{Analytic strong-deflection angle in the Schwarzschild limit}
\label{app:schwarzschild}

In this appendix, we present the analytic result for the strong-deflection angle in the Schwarzschild limit. When $\ell\rightarrow0$, the Dymnikova background reduces to the Schwarzschild spacetime, and the strong-deflection angle can be obtained analytically. We derive the corresponding strong-deflection coefficients and present the resulting expression for the deflection angle, which agrees with the analytic result obtained in Ref.~\cite{Bozza:2002zj,Kanai:2026xpw,Kanai:2026bag}.

In the Schwarzschild case, the functions $R(z,r_{\rm ph})$ and $f(z,r_{\rm ph})$ appearing in the integral expression for the deflection angle can be obtained analytically as
\begin{align}
R(z,r_{\rm ph})=&2-\frac{8}{9}s\gamma(1-2(1-z)^3),\\
f(z,r_{\rm ph})=&\dfrac{1}{\sqrt{z^2-\frac{2z^3}{3}}}+\dfrac{4s\gamma z(3-3z+z^2)}{9\sqrt{z^2-\frac{2z^3}{3}}},
\end{align}
where $z$ is defined in terms of the radial coordinate $r$ as in Eq.~\eqref{zcoor}.

Using these expressions, the regular contribution $b_R$ can be evaluated analytically and is given by
\begin{align}
b_R=-4\arctan\left(\frac{1}{\sqrt{3}}\right)+\log36-\frac{8s\gamma}{45}(18-4\sqrt{3}-5\log12+10\log(1+\sqrt{3})).
\end{align}
Combining the divergent and regular contributions, the strong-deflection coefficient $\bar b$ is obtained as
\begin{align}
\bar{b}=-\pi-4\arctan\left(\frac{1}{\sqrt{3}}\right)+\log1296+\frac{8}{45}s\gamma\left(-8+4\sqrt{3}+\log1934917632-10\log(1+\sqrt{3})\right).
\end{align}
For the nonzero values of $\ell$, the corresponding coefficient $\bar b$ is obtained numerically by evaluating the divergent and regular contributions, as summarized in Table~\ref{tab:b}.

Finally, in the Schwarzschild spacetime, the strong-deflection angle can be expressed in terms of the angular position $\theta$ as
\begin{align}
\alpha(\theta)=&-\left(1+\frac{4s\gamma}{9}\right)\log\left(\frac{\theta D_{OL}}{3\sqrt{3}-\frac{4s\gamma}{\sqrt{3}}}-1 \right)+\bar{b}+\mathcal{O}\left((u-u_{\rm ph})\log(u-u_{\rm ph})\right).
\end{align}
For both the Schwarzschild limit, $\ell\rightarrow0$, and the nonzero values of $\ell$, the strong-deflection quantities $u_{\rm ph}$, $\bar a$, $b_R$, and $\bar b$ are summarized in Tables~\ref{tab:bara}, \ref{tab:uph} and \ref{tab:b}, with the $\ell\rightarrow0$ case given analytically and the nonzero $\ell$ cases evaluated numerically.

\bibliography{references}

\begin{thebibliography}{70}%
\makeatletter
\providecommand \@ifxundefined [1]{%
 \@ifx{#1\undefined}
}%
\providecommand \@ifnum [1]{%
 \ifnum #1\expandafter \@firstoftwo
 \else \expandafter \@secondoftwo
 \fi
}%
\providecommand \@ifx [1]{%
 \ifx #1\expandafter \@firstoftwo
 \else \expandafter \@secondoftwo
 \fi
}%
\providecommand \natexlab [1]{#1}%
\providecommand \enquote  [1]{``#1''}%
\providecommand \bibnamefont  [1]{#1}%
\providecommand \bibfnamefont [1]{#1}%
\providecommand \citenamefont [1]{#1}%
\providecommand \href@noop [0]{\@secondoftwo}%
\providecommand \href [0]{\begingroup \@sanitize@url \@href}%
\providecommand \@href[1]{\@@startlink{#1}\@@href}%
\providecommand \@@href[1]{\endgroup#1\@@endlink}%
\providecommand \@sanitize@url [0]{\catcode `\\12\catcode `\$12\catcode
  `\&12\catcode `\#12\catcode `\^12\catcode `\_12\catcode `\%12\relax}%
\providecommand \@@startlink[1]{}%
\providecommand \@@endlink[0]{}%
\providecommand \url  [0]{\begingroup\@sanitize@url \@url }%
\providecommand \@url [1]{\endgroup\@href {#1}{\urlprefix }}%
\providecommand \urlprefix  [0]{URL }%
\providecommand \Eprint [0]{\href }%
\providecommand \doibase [0]{https://doi.org/}%
\providecommand \selectlanguage [0]{\@gobble}%
\providecommand \bibinfo  [0]{\@secondoftwo}%
\providecommand \bibfield  [0]{\@secondoftwo}%
\providecommand \translation [1]{[#1]}%
\providecommand \BibitemOpen [0]{}%
\providecommand \bibitemStop [0]{}%
\providecommand \bibitemNoStop [0]{.\EOS\space}%
\providecommand \EOS [0]{\spacefactor3000\relax}%
\providecommand \BibitemShut  [1]{\csname bibitem#1\endcsname}%
\let\auto@bib@innerbib\@empty
\bibitem [{\citenamefont {Falcke}\ \emph {et~al.}(2000)\citenamefont {Falcke},
  \citenamefont {Melia},\ and\ \citenamefont {Agol}}]{Falcke:1999pj}%
  \BibitemOpen
  \bibfield  {author} {\bibinfo {author} {\bibfnamefont {H.}~\bibnamefont
  {Falcke}}, \bibinfo {author} {\bibfnamefont {F.}~\bibnamefont {Melia}},\ and\
  \bibinfo {author} {\bibfnamefont {E.}~\bibnamefont {Agol}},\ }\bibfield
  {title} {\bibinfo {title} {{Viewing the shadow of the black hole at the
  galactic center}},\ }\href {https://doi.org/10.1086/312423} {\bibfield
  {journal} {\bibinfo  {journal} {Astrophys. J. Lett.}\ }\textbf {\bibinfo
  {volume} {528}},\ \bibinfo {pages} {L13} (\bibinfo {year} {2000})},\ \Eprint
  {https://arxiv.org/abs/astro-ph/9912263} {arXiv:astro-ph/9912263}
  \BibitemShut {NoStop}%
\bibitem [{\citenamefont {Akiyama}\ \emph {et~al.}(2019)\citenamefont {Akiyama}
  \emph {et~al.}}]{EventHorizonTelescope:2019dse}%
  \BibitemOpen
  \bibfield  {author} {\bibinfo {author} {\bibfnamefont {K.}~\bibnamefont
  {Akiyama}} \emph {et~al.} (\bibinfo {collaboration} {Event Horizon
  Telescope}),\ }\bibfield  {title} {\bibinfo {title} {{First M87 Event Horizon
  Telescope Results. I. The Shadow of the Supermassive Black Hole}},\ }\href
  {https://doi.org/10.3847/2041-8213/ab0ec7} {\bibfield  {journal} {\bibinfo
  {journal} {Astrophys. J. Lett.}\ }\textbf {\bibinfo {volume} {875}},\
  \bibinfo {pages} {L1} (\bibinfo {year} {2019})},\ \Eprint
  {https://arxiv.org/abs/1906.11238} {arXiv:1906.11238 [astro-ph.GA]}
  \BibitemShut {NoStop}%
\bibitem [{\citenamefont {Akiyama}\ \emph {et~al.}(2022)\citenamefont {Akiyama}
  \emph {et~al.}}]{EventHorizonTelescope:2022wkp}%
  \BibitemOpen
  \bibfield  {author} {\bibinfo {author} {\bibfnamefont {K.}~\bibnamefont
  {Akiyama}} \emph {et~al.} (\bibinfo {collaboration} {Event Horizon
  Telescope}),\ }\bibfield  {title} {\bibinfo {title} {{First Sagittarius A*
  Event Horizon Telescope Results. I. The Shadow of the Supermassive Black Hole
  in the Center of the Milky Way}},\ }\href
  {https://doi.org/10.3847/2041-8213/ac6674} {\bibfield  {journal} {\bibinfo
  {journal} {Astrophys. J. Lett.}\ }\textbf {\bibinfo {volume} {930}},\
  \bibinfo {pages} {L12} (\bibinfo {year} {2022})},\ \Eprint
  {https://arxiv.org/abs/2311.08680} {arXiv:2311.08680 [astro-ph.HE]}
  \BibitemShut {NoStop}%
\bibitem [{\citenamefont {Vagnozzi}\ \emph {et~al.}(2023)\citenamefont
  {Vagnozzi} \emph {et~al.}}]{Vagnozzi:2022moj}%
  \BibitemOpen
  \bibfield  {author} {\bibinfo {author} {\bibfnamefont {S.}~\bibnamefont
  {Vagnozzi}} \emph {et~al.},\ }\bibfield  {title} {\bibinfo {title}
  {{Horizon-scale tests of gravity theories and fundamental physics from the
  Event Horizon Telescope image of Sagittarius A}},\ }\href
  {https://doi.org/10.1088/1361-6382/acd97b} {\bibfield  {journal} {\bibinfo
  {journal} {Class. Quant. Grav.}\ }\textbf {\bibinfo {volume} {40}},\ \bibinfo
  {pages} {165007} (\bibinfo {year} {2023})},\ \Eprint
  {https://arxiv.org/abs/2205.07787} {arXiv:2205.07787 [gr-qc]} \BibitemShut
  {NoStop}%
\bibitem [{\citenamefont {Virbhadra}\ and\ \citenamefont
  {Ellis}(2000)}]{Virbhadra:1999nm}%
  \BibitemOpen
  \bibfield  {author} {\bibinfo {author} {\bibfnamefont {K.~S.}\ \bibnamefont
  {Virbhadra}}\ and\ \bibinfo {author} {\bibfnamefont {G.~F.~R.}\ \bibnamefont
  {Ellis}},\ }\bibfield  {title} {\bibinfo {title} {{Schwarzschild black hole
  lensing}},\ }\href {https://doi.org/10.1103/PhysRevD.62.084003} {\bibfield
  {journal} {\bibinfo  {journal} {Phys. Rev. D}\ }\textbf {\bibinfo {volume}
  {62}},\ \bibinfo {pages} {084003} (\bibinfo {year} {2000})},\ \Eprint
  {https://arxiv.org/abs/astro-ph/9904193} {arXiv:astro-ph/9904193}
  \BibitemShut {NoStop}%
\bibitem [{\citenamefont {Bozza}\ \emph {et~al.}(2001)\citenamefont {Bozza},
  \citenamefont {Capozziello}, \citenamefont {Iovane},\ and\ \citenamefont
  {Scarpetta}}]{Bozza:2001xd}%
  \BibitemOpen
  \bibfield  {author} {\bibinfo {author} {\bibfnamefont {V.}~\bibnamefont
  {Bozza}}, \bibinfo {author} {\bibfnamefont {S.}~\bibnamefont {Capozziello}},
  \bibinfo {author} {\bibfnamefont {G.}~\bibnamefont {Iovane}},\ and\ \bibinfo
  {author} {\bibfnamefont {G.}~\bibnamefont {Scarpetta}},\ }\bibfield  {title}
  {\bibinfo {title} {{Strong field limit of black hole gravitational
  lensing}},\ }\href {https://doi.org/10.1023/A:1012292927358} {\bibfield
  {journal} {\bibinfo  {journal} {Gen. Rel. Grav.}\ }\textbf {\bibinfo {volume}
  {33}},\ \bibinfo {pages} {1535} (\bibinfo {year} {2001})},\ \Eprint
  {https://arxiv.org/abs/gr-qc/0102068} {arXiv:gr-qc/0102068} \BibitemShut
  {NoStop}%
\bibitem [{\citenamefont {Bozza}(2002)}]{Bozza:2002zj}%
  \BibitemOpen
  \bibfield  {author} {\bibinfo {author} {\bibfnamefont {V.}~\bibnamefont
  {Bozza}},\ }\bibfield  {title} {\bibinfo {title} {{Gravitational lensing in
  the strong field limit}},\ }\href
  {https://doi.org/10.1103/PhysRevD.66.103001} {\bibfield  {journal} {\bibinfo
  {journal} {Phys. Rev. D}\ }\textbf {\bibinfo {volume} {66}},\ \bibinfo
  {pages} {103001} (\bibinfo {year} {2002})},\ \Eprint
  {https://arxiv.org/abs/gr-qc/0208075} {arXiv:gr-qc/0208075} \BibitemShut
  {NoStop}%
\bibitem [{\citenamefont {Gibbons}\ and\ \citenamefont
  {Werner}(2008)}]{Gibbons:2008rj}%
  \BibitemOpen
  \bibfield  {author} {\bibinfo {author} {\bibfnamefont {G.~W.}\ \bibnamefont
  {Gibbons}}\ and\ \bibinfo {author} {\bibfnamefont {M.~C.}\ \bibnamefont
  {Werner}},\ }\bibfield  {title} {\bibinfo {title} {{Applications of the
  Gauss-Bonnet theorem to gravitational lensing}},\ }\href
  {https://doi.org/10.1088/0264-9381/25/23/235009} {\bibfield  {journal}
  {\bibinfo  {journal} {Class. Quant. Grav.}\ }\textbf {\bibinfo {volume}
  {25}},\ \bibinfo {pages} {235009} (\bibinfo {year} {2008})},\ \Eprint
  {https://arxiv.org/abs/0807.0854} {arXiv:0807.0854 [gr-qc]} \BibitemShut
  {NoStop}%
\bibitem [{\citenamefont {Bozza}(2003)}]{Bozza:2002af}%
  \BibitemOpen
  \bibfield  {author} {\bibinfo {author} {\bibfnamefont {V.}~\bibnamefont
  {Bozza}},\ }\bibfield  {title} {\bibinfo {title} {{Quasiequatorial
  gravitational lensing by spinning black holes in the strong field limit}},\
  }\href {https://doi.org/10.1103/PhysRevD.67.103006} {\bibfield  {journal}
  {\bibinfo  {journal} {Phys. Rev. D}\ }\textbf {\bibinfo {volume} {67}},\
  \bibinfo {pages} {103006} (\bibinfo {year} {2003})},\ \Eprint
  {https://arxiv.org/abs/gr-qc/0210109} {arXiv:gr-qc/0210109} \BibitemShut
  {NoStop}%
\bibitem [{\citenamefont {Bozza}\ and\ \citenamefont
  {Mancini}(2004)}]{Bozza:2003cp}%
  \BibitemOpen
  \bibfield  {author} {\bibinfo {author} {\bibfnamefont {V.}~\bibnamefont
  {Bozza}}\ and\ \bibinfo {author} {\bibfnamefont {L.}~\bibnamefont
  {Mancini}},\ }\bibfield  {title} {\bibinfo {title} {{Time delay in black hole
  gravitational lensing as a distance estimator}},\ }\href
  {https://doi.org/10.1023/B:GERG.0000010486.58026.4f} {\bibfield  {journal}
  {\bibinfo  {journal} {Gen. Rel. Grav.}\ }\textbf {\bibinfo {volume} {36}},\
  \bibinfo {pages} {435} (\bibinfo {year} {2004})},\ \Eprint
  {https://arxiv.org/abs/gr-qc/0305007} {arXiv:gr-qc/0305007} \BibitemShut
  {NoStop}%
\bibitem [{\citenamefont {{Bardeen}}(1968)}]{1968qtr..conf...87B}%
  \BibitemOpen
  \bibfield  {author} {\bibinfo {author} {\bibfnamefont {J.}~\bibnamefont
  {{Bardeen}}},\ }\bibfield  {title} {\bibinfo {title} {{Non-singular general
  relativistic gravitational collapse}},\ }in\ \href@noop {} {\emph {\bibinfo
  {booktitle} {Proceedings of the 5th International Conference on Gravitation
  and the Theory of Relativity}}}\ (\bibinfo {year} {1968})\ p.~\bibinfo
  {pages} {87}\BibitemShut {NoStop}%
\bibitem [{\citenamefont {Ayon-Beato}\ and\ \citenamefont
  {Garcia}(1998)}]{Ayon-Beato:1998hmi}%
  \BibitemOpen
  \bibfield  {author} {\bibinfo {author} {\bibfnamefont {E.}~\bibnamefont
  {Ayon-Beato}}\ and\ \bibinfo {author} {\bibfnamefont {A.}~\bibnamefont
  {Garcia}},\ }\bibfield  {title} {\bibinfo {title} {{Regular black hole in
  general relativity coupled to nonlinear electrodynamics}},\ }\href
  {https://doi.org/10.1103/PhysRevLett.80.5056} {\bibfield  {journal} {\bibinfo
   {journal} {Phys. Rev. Lett.}\ }\textbf {\bibinfo {volume} {80}},\ \bibinfo
  {pages} {5056} (\bibinfo {year} {1998})},\ \Eprint
  {https://arxiv.org/abs/gr-qc/9911046} {arXiv:gr-qc/9911046} \BibitemShut
  {NoStop}%
\bibitem [{\citenamefont {Ayon-Beato}\ and\ \citenamefont
  {Garcia}(1999)}]{Ayon-Beato:1999kuh}%
  \BibitemOpen
  \bibfield  {author} {\bibinfo {author} {\bibfnamefont {E.}~\bibnamefont
  {Ayon-Beato}}\ and\ \bibinfo {author} {\bibfnamefont {A.}~\bibnamefont
  {Garcia}},\ }\bibfield  {title} {\bibinfo {title} {{New regular black hole
  solution from nonlinear electrodynamics}},\ }\href
  {https://doi.org/10.1016/S0370-2693(99)01038-2} {\bibfield  {journal}
  {\bibinfo  {journal} {Phys. Lett. B}\ }\textbf {\bibinfo {volume} {464}},\
  \bibinfo {pages} {25} (\bibinfo {year} {1999})},\ \Eprint
  {https://arxiv.org/abs/hep-th/9911174} {arXiv:hep-th/9911174} \BibitemShut
  {NoStop}%
\bibitem [{\citenamefont {Hayward}(2006)}]{Hayward:2005gi}%
  \BibitemOpen
  \bibfield  {author} {\bibinfo {author} {\bibfnamefont {S.~A.}\ \bibnamefont
  {Hayward}},\ }\bibfield  {title} {\bibinfo {title} {{Formation and
  evaporation of regular black holes}},\ }\href
  {https://doi.org/10.1103/PhysRevLett.96.031103} {\bibfield  {journal}
  {\bibinfo  {journal} {Phys. Rev. Lett.}\ }\textbf {\bibinfo {volume} {96}},\
  \bibinfo {pages} {031103} (\bibinfo {year} {2006})},\ \Eprint
  {https://arxiv.org/abs/gr-qc/0506126} {arXiv:gr-qc/0506126} \BibitemShut
  {NoStop}%
\bibitem [{\citenamefont {Ashtekar}\ and\ \citenamefont
  {Bojowald}(2006)}]{Ashtekar:2005qt}%
  \BibitemOpen
  \bibfield  {author} {\bibinfo {author} {\bibfnamefont {A.}~\bibnamefont
  {Ashtekar}}\ and\ \bibinfo {author} {\bibfnamefont {M.}~\bibnamefont
  {Bojowald}},\ }\bibfield  {title} {\bibinfo {title} {{Quantum geometry and
  the Schwarzschild singularity}},\ }\href
  {https://doi.org/10.1088/0264-9381/23/2/008} {\bibfield  {journal} {\bibinfo
  {journal} {Class. Quant. Grav.}\ }\textbf {\bibinfo {volume} {23}},\ \bibinfo
  {pages} {391} (\bibinfo {year} {2006})},\ \Eprint
  {https://arxiv.org/abs/gr-qc/0509075} {arXiv:gr-qc/0509075} \BibitemShut
  {NoStop}%
\bibitem [{\citenamefont {Modesto}(2006)}]{Modesto:2005zm}%
  \BibitemOpen
  \bibfield  {author} {\bibinfo {author} {\bibfnamefont {L.}~\bibnamefont
  {Modesto}},\ }\bibfield  {title} {\bibinfo {title} {{Loop quantum black
  hole}},\ }\href {https://doi.org/10.1088/0264-9381/23/18/006} {\bibfield
  {journal} {\bibinfo  {journal} {Class. Quant. Grav.}\ }\textbf {\bibinfo
  {volume} {23}},\ \bibinfo {pages} {5587} (\bibinfo {year} {2006})},\ \Eprint
  {https://arxiv.org/abs/gr-qc/0509078} {arXiv:gr-qc/0509078} \BibitemShut
  {NoStop}%
\bibitem [{\citenamefont {Weinberg}(1979)}]{Weinberg:1978kz}%
  \BibitemOpen
  \bibfield  {author} {\bibinfo {author} {\bibfnamefont {S.}~\bibnamefont
  {Weinberg}},\ }\bibfield  {title} {\bibinfo {title} {{Phenomenological
  Lagrangians}},\ }\href {https://doi.org/10.1016/0378-4371(79)90223-1}
  {\bibfield  {journal} {\bibinfo  {journal} {Physica A}\ }\textbf {\bibinfo
  {volume} {96}},\ \bibinfo {pages} {327} (\bibinfo {year} {1979})}\BibitemShut
  {NoStop}%
\bibitem [{\citenamefont {Donoghue}(1994{\natexlab{a}})}]{Donoghue:1993eb}%
  \BibitemOpen
  \bibfield  {author} {\bibinfo {author} {\bibfnamefont {J.~F.}\ \bibnamefont
  {Donoghue}},\ }\bibfield  {title} {\bibinfo {title} {{Leading quantum
  correction to the Newtonian potential}},\ }\href
  {https://doi.org/10.1103/PhysRevLett.72.2996} {\bibfield  {journal} {\bibinfo
   {journal} {Phys. Rev. Lett.}\ }\textbf {\bibinfo {volume} {72}},\ \bibinfo
  {pages} {2996} (\bibinfo {year} {1994}{\natexlab{a}})},\ \Eprint
  {https://arxiv.org/abs/gr-qc/9310024} {arXiv:gr-qc/9310024} \BibitemShut
  {NoStop}%
\bibitem [{\citenamefont {Donoghue}(1994{\natexlab{b}})}]{Donoghue:1994dn}%
  \BibitemOpen
  \bibfield  {author} {\bibinfo {author} {\bibfnamefont {J.~F.}\ \bibnamefont
  {Donoghue}},\ }\bibfield  {title} {\bibinfo {title} {{General relativity as
  an effective field theory: The leading quantum corrections}},\ }\href
  {https://doi.org/10.1103/PhysRevD.50.3874} {\bibfield  {journal} {\bibinfo
  {journal} {Phys. Rev. D}\ }\textbf {\bibinfo {volume} {50}},\ \bibinfo
  {pages} {3874} (\bibinfo {year} {1994}{\natexlab{b}})},\ \Eprint
  {https://arxiv.org/abs/gr-qc/9405057} {arXiv:gr-qc/9405057} \BibitemShut
  {NoStop}%
\bibitem [{\citenamefont {Burgess}(2004)}]{Burgess:2003jk}%
  \BibitemOpen
  \bibfield  {author} {\bibinfo {author} {\bibfnamefont {C.~P.}\ \bibnamefont
  {Burgess}},\ }\bibfield  {title} {\bibinfo {title} {{Quantum gravity in
  everyday life: General relativity as an effective field theory}},\ }\href
  {https://doi.org/10.12942/lrr-2004-5} {\bibfield  {journal} {\bibinfo
  {journal} {Living Rev. Rel.}\ }\textbf {\bibinfo {volume} {7}},\ \bibinfo
  {pages} {5} (\bibinfo {year} {2004})},\ \Eprint
  {https://arxiv.org/abs/gr-qc/0311082} {arXiv:gr-qc/0311082} \BibitemShut
  {NoStop}%
\bibitem [{\citenamefont {Scharnhorst}(1990)}]{Scharnhorst:1990sr}%
  \BibitemOpen
  \bibfield  {author} {\bibinfo {author} {\bibfnamefont {K.}~\bibnamefont
  {Scharnhorst}},\ }\bibfield  {title} {\bibinfo {title} {{On propagation of
  light in the vacuum between plates}},\ }\href
  {https://doi.org/10.1016/0370-2693(90)90997-K} {\bibfield  {journal}
  {\bibinfo  {journal} {Phys. Lett. B}\ }\textbf {\bibinfo {volume} {236}},\
  \bibinfo {pages} {354} (\bibinfo {year} {1990})},\ \bibinfo {note} {[Erratum:
  Phys.Lett.B 787, 204--204 (2018)]}\BibitemShut {NoStop}%
\bibitem [{\citenamefont {Barton}(1990)}]{Barton:1989dq}%
  \BibitemOpen
  \bibfield  {author} {\bibinfo {author} {\bibfnamefont {G.}~\bibnamefont
  {Barton}},\ }\bibfield  {title} {\bibinfo {title} {{Faster Than $c$ Light
  Between Parallel Mirrors: The Scharnhorst Effect Rederived}},\ }\href
  {https://doi.org/10.1016/0370-2693(90)91224-Y} {\bibfield  {journal}
  {\bibinfo  {journal} {Phys. Lett. B}\ }\textbf {\bibinfo {volume} {237}},\
  \bibinfo {pages} {559} (\bibinfo {year} {1990})}\BibitemShut {NoStop}%
\bibitem [{\citenamefont {Barton}\ and\ \citenamefont
  {Scharnhorst}(1993)}]{Barton:1992pq}%
  \BibitemOpen
  \bibfield  {author} {\bibinfo {author} {\bibfnamefont {G.}~\bibnamefont
  {Barton}}\ and\ \bibinfo {author} {\bibfnamefont {K.}~\bibnamefont
  {Scharnhorst}},\ }\bibfield  {title} {\bibinfo {title} {{QED between parallel
  mirrors: Light signals faster than c, or amplified by the vacuum}},\ }\href
  {https://doi.org/10.1088/0305-4470/26/8/024} {\bibfield  {journal} {\bibinfo
  {journal} {J. Phys. A}\ }\textbf {\bibinfo {volume} {26}},\ \bibinfo {pages}
  {2037} (\bibinfo {year} {1993})}\BibitemShut {NoStop}%
\bibitem [{\citenamefont {Latorre}\ \emph {et~al.}(1995)\citenamefont
  {Latorre}, \citenamefont {Pascual},\ and\ \citenamefont
  {Tarrach}}]{Latorre:1994cv}%
  \BibitemOpen
  \bibfield  {author} {\bibinfo {author} {\bibfnamefont {J.~I.}\ \bibnamefont
  {Latorre}}, \bibinfo {author} {\bibfnamefont {P.}~\bibnamefont {Pascual}},\
  and\ \bibinfo {author} {\bibfnamefont {R.}~\bibnamefont {Tarrach}},\
  }\bibfield  {title} {\bibinfo {title} {{Speed of light in nontrivial
  vacua}},\ }\href {https://doi.org/10.1016/0550-3213(94)00490-6} {\bibfield
  {journal} {\bibinfo  {journal} {Nucl. Phys. B}\ }\textbf {\bibinfo {volume}
  {437}},\ \bibinfo {pages} {60} (\bibinfo {year} {1995})},\ \Eprint
  {https://arxiv.org/abs/hep-th/9408016} {arXiv:hep-th/9408016} \BibitemShut
  {NoStop}%
\bibitem [{\citenamefont {Dittrich}\ and\ \citenamefont
  {Gies}(1998)}]{Dittrich:1998fy}%
  \BibitemOpen
  \bibfield  {author} {\bibinfo {author} {\bibfnamefont {W.}~\bibnamefont
  {Dittrich}}\ and\ \bibinfo {author} {\bibfnamefont {H.}~\bibnamefont
  {Gies}},\ }\bibfield  {title} {\bibinfo {title} {{Light propagation in
  nontrivial QED vacua}},\ }\href {https://doi.org/10.1103/PhysRevD.58.025004}
  {\bibfield  {journal} {\bibinfo  {journal} {Phys. Rev. D}\ }\textbf {\bibinfo
  {volume} {58}},\ \bibinfo {pages} {025004} (\bibinfo {year} {1998})},\
  \Eprint {https://arxiv.org/abs/hep-ph/9804375} {arXiv:hep-ph/9804375}
  \BibitemShut {NoStop}%
\bibitem [{\citenamefont {De~Lorenci}\ \emph {et~al.}(2000)\citenamefont
  {De~Lorenci}, \citenamefont {Klippert}, \citenamefont {Novello},\ and\
  \citenamefont {Salim}}]{DeLorenci:2000yh}%
  \BibitemOpen
  \bibfield  {author} {\bibinfo {author} {\bibfnamefont {V.~A.}\ \bibnamefont
  {De~Lorenci}}, \bibinfo {author} {\bibfnamefont {R.}~\bibnamefont
  {Klippert}}, \bibinfo {author} {\bibfnamefont {M.}~\bibnamefont {Novello}},\
  and\ \bibinfo {author} {\bibfnamefont {J.~M.}\ \bibnamefont {Salim}},\
  }\bibfield  {title} {\bibinfo {title} {{Light propagation in nonlinear
  electrodynamics}},\ }\href {https://doi.org/10.1016/S0370-2693(00)00522-0}
  {\bibfield  {journal} {\bibinfo  {journal} {Phys. Lett. B}\ }\textbf
  {\bibinfo {volume} {482}},\ \bibinfo {pages} {134} (\bibinfo {year}
  {2000})},\ \Eprint {https://arxiv.org/abs/gr-qc/0005049}
  {arXiv:gr-qc/0005049} \BibitemShut {NoStop}%
\bibitem [{\citenamefont {Drummond}\ and\ \citenamefont
  {Hathrell}(1980)}]{Drummond:1979pp}%
  \BibitemOpen
  \bibfield  {author} {\bibinfo {author} {\bibfnamefont {I.~T.}\ \bibnamefont
  {Drummond}}\ and\ \bibinfo {author} {\bibfnamefont {S.~J.}\ \bibnamefont
  {Hathrell}},\ }\bibfield  {title} {\bibinfo {title} {{QED Vacuum Polarization
  in a Background Gravitational Field and Its Effect on the Velocity of
  Photons}},\ }\href {https://doi.org/10.1103/PhysRevD.22.343} {\bibfield
  {journal} {\bibinfo  {journal} {Phys. Rev. D}\ }\textbf {\bibinfo {volume}
  {22}},\ \bibinfo {pages} {343} (\bibinfo {year} {1980})}\BibitemShut
  {NoStop}%
\bibitem [{\citenamefont {Daniels}\ and\ \citenamefont
  {Shore}(1994)}]{Daniels:1993yi}%
  \BibitemOpen
  \bibfield  {author} {\bibinfo {author} {\bibfnamefont {R.~D.}\ \bibnamefont
  {Daniels}}\ and\ \bibinfo {author} {\bibfnamefont {G.~M.}\ \bibnamefont
  {Shore}},\ }\bibfield  {title} {\bibinfo {title} {{'Faster than light'
  photons and charged black holes}},\ }\href
  {https://doi.org/10.1016/0550-3213(94)90291-7} {\bibfield  {journal}
  {\bibinfo  {journal} {Nucl. Phys. B}\ }\textbf {\bibinfo {volume} {425}},\
  \bibinfo {pages} {634} (\bibinfo {year} {1994})},\ \Eprint
  {https://arxiv.org/abs/hep-th/9310114} {arXiv:hep-th/9310114} \BibitemShut
  {NoStop}%
\bibitem [{\citenamefont {Shore}(1996)}]{Shore:1995fz}%
  \BibitemOpen
  \bibfield  {author} {\bibinfo {author} {\bibfnamefont {G.~M.}\ \bibnamefont
  {Shore}},\ }\bibfield  {title} {\bibinfo {title} {{'Faster than light'
  photons in gravitational fields: Causality, anomalies and horizons}},\ }\href
  {https://doi.org/10.1016/0550-3213(95)00646-X} {\bibfield  {journal}
  {\bibinfo  {journal} {Nucl. Phys. B}\ }\textbf {\bibinfo {volume} {460}},\
  \bibinfo {pages} {379} (\bibinfo {year} {1996})},\ \Eprint
  {https://arxiv.org/abs/gr-qc/9504041} {arXiv:gr-qc/9504041} \BibitemShut
  {NoStop}%
\bibitem [{\citenamefont {Daniels}\ and\ \citenamefont
  {Shore}(1996)}]{Daniels:1995yw}%
  \BibitemOpen
  \bibfield  {author} {\bibinfo {author} {\bibfnamefont {R.~D.}\ \bibnamefont
  {Daniels}}\ and\ \bibinfo {author} {\bibfnamefont {G.~M.}\ \bibnamefont
  {Shore}},\ }\bibfield  {title} {\bibinfo {title} {{'Faster than light'
  photons and rotating black holes}},\ }\href
  {https://doi.org/10.1016/0370-2693(95)01468-3} {\bibfield  {journal}
  {\bibinfo  {journal} {Phys. Lett. B}\ }\textbf {\bibinfo {volume} {367}},\
  \bibinfo {pages} {75} (\bibinfo {year} {1996})},\ \Eprint
  {https://arxiv.org/abs/gr-qc/9508048} {arXiv:gr-qc/9508048} \BibitemShut
  {NoStop}%
\bibitem [{\citenamefont {Shore}(2007)}]{Shore:2007um}%
  \BibitemOpen
  \bibfield  {author} {\bibinfo {author} {\bibfnamefont {G.~M.}\ \bibnamefont
  {Shore}},\ }\bibfield  {title} {\bibinfo {title} {{Superluminality and UV
  completion}},\ }\href {https://doi.org/10.1016/j.nuclphysb.2007.03.034}
  {\bibfield  {journal} {\bibinfo  {journal} {Nucl. Phys. B}\ }\textbf
  {\bibinfo {volume} {778}},\ \bibinfo {pages} {219} (\bibinfo {year}
  {2007})},\ \Eprint {https://arxiv.org/abs/hep-th/0701185}
  {arXiv:hep-th/0701185} \BibitemShut {NoStop}%
\bibitem [{\citenamefont {Cho}(1997)}]{Cho:1997vg}%
  \BibitemOpen
  \bibfield  {author} {\bibinfo {author} {\bibfnamefont {H.~T.}\ \bibnamefont
  {Cho}},\ }\bibfield  {title} {\bibinfo {title} {{'Faster than light' photons
  in dilaton black hole space-times}},\ }\href
  {https://doi.org/10.1103/PhysRevD.56.6416} {\bibfield  {journal} {\bibinfo
  {journal} {Phys. Rev. D}\ }\textbf {\bibinfo {volume} {56}},\ \bibinfo
  {pages} {6416} (\bibinfo {year} {1997})},\ \Eprint
  {https://arxiv.org/abs/gr-qc/9704014} {arXiv:gr-qc/9704014} \BibitemShut
  {NoStop}%
\bibitem [{\citenamefont {Izumi}(2014)}]{Izumi:2014loa}%
  \BibitemOpen
  \bibfield  {author} {\bibinfo {author} {\bibfnamefont {K.}~\bibnamefont
  {Izumi}},\ }\bibfield  {title} {\bibinfo {title} {{Causal Structures in
  Gauss-Bonnet gravity}},\ }\href {https://doi.org/10.1103/PhysRevD.90.044037}
  {\bibfield  {journal} {\bibinfo  {journal} {Phys. Rev. D}\ }\textbf {\bibinfo
  {volume} {90}},\ \bibinfo {pages} {044037} (\bibinfo {year} {2014})},\
  \Eprint {https://arxiv.org/abs/1406.0677} {arXiv:1406.0677 [gr-qc]}
  \BibitemShut {NoStop}%
\bibitem [{\citenamefont {Reall}\ \emph {et~al.}(2014)\citenamefont {Reall},
  \citenamefont {Tanahashi},\ and\ \citenamefont {Way}}]{Reall:2014pwa}%
  \BibitemOpen
  \bibfield  {author} {\bibinfo {author} {\bibfnamefont {H.}~\bibnamefont
  {Reall}}, \bibinfo {author} {\bibfnamefont {N.}~\bibnamefont {Tanahashi}},\
  and\ \bibinfo {author} {\bibfnamefont {B.}~\bibnamefont {Way}},\ }\bibfield
  {title} {\bibinfo {title} {{Causality and Hyperbolicity of Lovelock
  Theories}},\ }\href {https://doi.org/10.1088/0264-9381/31/20/205005}
  {\bibfield  {journal} {\bibinfo  {journal} {Class. Quant. Grav.}\ }\textbf
  {\bibinfo {volume} {31}},\ \bibinfo {pages} {205005} (\bibinfo {year}
  {2014})},\ \Eprint {https://arxiv.org/abs/1406.3379} {arXiv:1406.3379
  [hep-th]} \BibitemShut {NoStop}%
\bibitem [{\citenamefont {Allahyari}\ \emph {et~al.}(2020)\citenamefont
  {Allahyari}, \citenamefont {Khodadi}, \citenamefont {Vagnozzi},\ and\
  \citenamefont {Mota}}]{Allahyari:2019jqz}%
  \BibitemOpen
  \bibfield  {author} {\bibinfo {author} {\bibfnamefont {A.}~\bibnamefont
  {Allahyari}}, \bibinfo {author} {\bibfnamefont {M.}~\bibnamefont {Khodadi}},
  \bibinfo {author} {\bibfnamefont {S.}~\bibnamefont {Vagnozzi}},\ and\
  \bibinfo {author} {\bibfnamefont {D.~F.}\ \bibnamefont {Mota}},\ }\bibfield
  {title} {\bibinfo {title} {{Magnetically charged black holes from non-linear
  electrodynamics and the Event Horizon Telescope}},\ }\href
  {https://doi.org/10.1088/1475-7516/2020/02/003} {\bibfield  {journal}
  {\bibinfo  {journal} {JCAP}\ }\textbf {\bibinfo {volume} {02}},\ \bibinfo
  {pages} {003}},\ \Eprint {https://arxiv.org/abs/1912.08231} {arXiv:1912.08231
  [gr-qc]} \BibitemShut {NoStop}%
\bibitem [{\citenamefont {Cao}\ and\ \citenamefont {Wu}(2021)}]{Cao:2021sty}%
  \BibitemOpen
  \bibfield  {author} {\bibinfo {author} {\bibfnamefont {L.-M.}\ \bibnamefont
  {Cao}}\ and\ \bibinfo {author} {\bibfnamefont {L.-B.}\ \bibnamefont {Wu}},\
  }\bibfield  {title} {\bibinfo {title} {{Hyperbolicity and Causality of
  Einstein-Gauss-Bonnet Gravity in Warped Product Spacetimes}},\ }\href
  {https://doi.org/10.1103/PhysRevD.103.064054} {\bibfield  {journal} {\bibinfo
   {journal} {Phys. Rev. D}\ }\textbf {\bibinfo {volume} {103}},\ \bibinfo
  {pages} {064054} (\bibinfo {year} {2021})},\ \Eprint
  {https://arxiv.org/abs/2101.02461} {arXiv:2101.02461 [gr-qc]} \BibitemShut
  {NoStop}%
\bibitem [{\citenamefont {Reall}(2021)}]{Reall:2021voz}%
  \BibitemOpen
  \bibfield  {author} {\bibinfo {author} {\bibfnamefont {H.~S.}\ \bibnamefont
  {Reall}},\ }\bibfield  {title} {\bibinfo {title} {{Causality in gravitational
  theories with second order equations of motion}},\ }\href
  {https://doi.org/10.1103/PhysRevD.103.084027} {\bibfield  {journal} {\bibinfo
   {journal} {Phys. Rev. D}\ }\textbf {\bibinfo {volume} {103}},\ \bibinfo
  {pages} {084027} (\bibinfo {year} {2021})},\ \Eprint
  {https://arxiv.org/abs/2101.11623} {arXiv:2101.11623 [gr-qc]} \BibitemShut
  {NoStop}%
\bibitem [{\citenamefont {Davies}\ and\ \citenamefont
  {Reall}(2022)}]{Davies:2021frz}%
  \BibitemOpen
  \bibfield  {author} {\bibinfo {author} {\bibfnamefont {I.}~\bibnamefont
  {Davies}}\ and\ \bibinfo {author} {\bibfnamefont {H.~S.}\ \bibnamefont
  {Reall}},\ }\bibfield  {title} {\bibinfo {title} {{Well-posed formulation of
  Einstein-Maxwell effective field theory}},\ }\href
  {https://doi.org/10.1103/PhysRevD.106.104019} {\bibfield  {journal} {\bibinfo
   {journal} {Phys. Rev. D}\ }\textbf {\bibinfo {volume} {106}},\ \bibinfo
  {pages} {104019} (\bibinfo {year} {2022})},\ \Eprint
  {https://arxiv.org/abs/2112.05603} {arXiv:2112.05603 [gr-qc]} \BibitemShut
  {NoStop}%
\bibitem [{\citenamefont {Fu}\ \emph {et~al.}(2025)\citenamefont {Fu},
  \citenamefont {Izumi},\ and\ \citenamefont {Yoshida}}]{Fu:2025oxr}%
  \BibitemOpen
  \bibfield  {author} {\bibinfo {author} {\bibfnamefont {L.}~\bibnamefont
  {Fu}}, \bibinfo {author} {\bibfnamefont {K.}~\bibnamefont {Izumi}},\ and\
  \bibinfo {author} {\bibfnamefont {D.}~\bibnamefont {Yoshida}},\ }\bibfield
  {title} {\bibinfo {title} {{Consistency between bulk and boundary causalities
  in asymptotically anti-de Sitter spacetimes}},\ }\href
  {https://doi.org/10.1007/JHEP10(2025)184} {\bibfield  {journal} {\bibinfo
  {journal} {JHEP}\ }\textbf {\bibinfo {volume} {10}},\ \bibinfo {pages}
  {184}},\ \Eprint {https://arxiv.org/abs/2504.15910} {arXiv:2504.15910
  [gr-qc]} \BibitemShut {NoStop}%
\bibitem [{\citenamefont {Kanai}(2026{\natexlab{a}})}]{Kanai:2026ohg}%
  \BibitemOpen
  \bibfield  {author} {\bibinfo {author} {\bibfnamefont {T.}~\bibnamefont
  {Kanai}},\ }\bibfield  {title} {\bibinfo {title} {{EFT Corrections to Photon
  Propagation and Gravitational Lensing in an Ellis-Bronnikov Wormhole}},\
  }\href@noop {} {\  (\bibinfo {year} {2026}{\natexlab{a}})},\ \Eprint
  {https://arxiv.org/abs/2608.08526} {arXiv:2608.08526 [gr-qc]} \BibitemShut
  {NoStop}%
\bibitem [{\citenamefont {Bueno}\ \emph
  {et~al.}(2025{\natexlab{a}})\citenamefont {Bueno}, \citenamefont {Cano},\
  and\ \citenamefont {Hennigar}}]{Bueno:2024dgm}%
  \BibitemOpen
  \bibfield  {author} {\bibinfo {author} {\bibfnamefont {P.}~\bibnamefont
  {Bueno}}, \bibinfo {author} {\bibfnamefont {P.~A.}\ \bibnamefont {Cano}},\
  and\ \bibinfo {author} {\bibfnamefont {R.~A.}\ \bibnamefont {Hennigar}},\
  }\bibfield  {title} {\bibinfo {title} {{Regular black holes from pure
  gravity}},\ }\href {https://doi.org/10.1016/j.physletb.2025.139260}
  {\bibfield  {journal} {\bibinfo  {journal} {Phys. Lett. B}\ }\textbf
  {\bibinfo {volume} {861}},\ \bibinfo {pages} {139260} (\bibinfo {year}
  {2025}{\natexlab{a}})},\ \Eprint {https://arxiv.org/abs/2403.04827}
  {arXiv:2403.04827 [gr-qc]} \BibitemShut {NoStop}%
\bibitem [{\citenamefont {Oliva}\ and\ \citenamefont
  {Ray}(2010)}]{Oliva:2010eb}%
  \BibitemOpen
  \bibfield  {author} {\bibinfo {author} {\bibfnamefont {J.}~\bibnamefont
  {Oliva}}\ and\ \bibinfo {author} {\bibfnamefont {S.}~\bibnamefont {Ray}},\
  }\bibfield  {title} {\bibinfo {title} {{A new cubic theory of gravity in five
  dimensions: Black hole, Birkhoff's theorem and C-function}},\ }\href
  {https://doi.org/10.1088/0264-9381/27/22/225002} {\bibfield  {journal}
  {\bibinfo  {journal} {Class. Quant. Grav.}\ }\textbf {\bibinfo {volume}
  {27}},\ \bibinfo {pages} {225002} (\bibinfo {year} {2010})},\ \Eprint
  {https://arxiv.org/abs/1003.4773} {arXiv:1003.4773 [gr-qc]} \BibitemShut
  {NoStop}%
\bibitem [{\citenamefont {Myers}\ and\ \citenamefont
  {Robinson}(2010)}]{Myers:2010ru}%
  \BibitemOpen
  \bibfield  {author} {\bibinfo {author} {\bibfnamefont {R.~C.}\ \bibnamefont
  {Myers}}\ and\ \bibinfo {author} {\bibfnamefont {B.}~\bibnamefont
  {Robinson}},\ }\bibfield  {title} {\bibinfo {title} {{Black Holes in
  Quasi-topological Gravity}},\ }\href
  {https://doi.org/10.1007/JHEP08(2010)067} {\bibfield  {journal} {\bibinfo
  {journal} {JHEP}\ }\textbf {\bibinfo {volume} {08}},\ \bibinfo {pages}
  {067}},\ \Eprint {https://arxiv.org/abs/1003.5357} {arXiv:1003.5357 [gr-qc]}
  \BibitemShut {NoStop}%
\bibitem [{\citenamefont {Dehghani}\ \emph {et~al.}(2012)\citenamefont
  {Dehghani}, \citenamefont {Bazrafshan}, \citenamefont {Mann}, \citenamefont
  {Mehdizadeh}, \citenamefont {Ghanaatian},\ and\ \citenamefont
  {Vahidinia}}]{Dehghani:2011vu}%
  \BibitemOpen
  \bibfield  {author} {\bibinfo {author} {\bibfnamefont {M.~H.}\ \bibnamefont
  {Dehghani}}, \bibinfo {author} {\bibfnamefont {A.}~\bibnamefont
  {Bazrafshan}}, \bibinfo {author} {\bibfnamefont {R.~B.}\ \bibnamefont
  {Mann}}, \bibinfo {author} {\bibfnamefont {M.~R.}\ \bibnamefont
  {Mehdizadeh}}, \bibinfo {author} {\bibfnamefont {M.}~\bibnamefont
  {Ghanaatian}},\ and\ \bibinfo {author} {\bibfnamefont {M.~H.}\ \bibnamefont
  {Vahidinia}},\ }\bibfield  {title} {\bibinfo {title} {{Black Holes in Quartic
  Quasitopological Gravity}},\ }\href
  {https://doi.org/10.1103/PhysRevD.85.104009} {\bibfield  {journal} {\bibinfo
  {journal} {Phys. Rev. D}\ }\textbf {\bibinfo {volume} {85}},\ \bibinfo
  {pages} {104009} (\bibinfo {year} {2012})},\ \Eprint
  {https://arxiv.org/abs/1109.4708} {arXiv:1109.4708 [hep-th]} \BibitemShut
  {NoStop}%
\bibitem [{\citenamefont {Cisterna}\ \emph {et~al.}(2017)\citenamefont
  {Cisterna}, \citenamefont {Guajardo}, \citenamefont {Hassaine},\ and\
  \citenamefont {Oliva}}]{Cisterna:2017umf}%
  \BibitemOpen
  \bibfield  {author} {\bibinfo {author} {\bibfnamefont {A.}~\bibnamefont
  {Cisterna}}, \bibinfo {author} {\bibfnamefont {L.}~\bibnamefont {Guajardo}},
  \bibinfo {author} {\bibfnamefont {M.}~\bibnamefont {Hassaine}},\ and\
  \bibinfo {author} {\bibfnamefont {J.}~\bibnamefont {Oliva}},\ }\bibfield
  {title} {\bibinfo {title} {{Quintic quasi-topological gravity}},\ }\href
  {https://doi.org/10.1007/JHEP04(2017)066} {\bibfield  {journal} {\bibinfo
  {journal} {JHEP}\ }\textbf {\bibinfo {volume} {04}},\ \bibinfo {pages}
  {066}},\ \Eprint {https://arxiv.org/abs/1702.04676} {arXiv:1702.04676
  [hep-th]} \BibitemShut {NoStop}%
\bibitem [{\citenamefont {Bueno}\ \emph {et~al.}(2019)\citenamefont {Bueno},
  \citenamefont {Cano}, \citenamefont {Moreno},\ and\ \citenamefont
  {Murcia}}]{Bueno:2019ltp}%
  \BibitemOpen
  \bibfield  {author} {\bibinfo {author} {\bibfnamefont {P.}~\bibnamefont
  {Bueno}}, \bibinfo {author} {\bibfnamefont {P.~A.}\ \bibnamefont {Cano}},
  \bibinfo {author} {\bibfnamefont {J.}~\bibnamefont {Moreno}},\ and\ \bibinfo
  {author} {\bibfnamefont {{\'A}.}~\bibnamefont {Murcia}},\ }\bibfield  {title}
  {\bibinfo {title} {{All higher-curvature gravities as Generalized
  quasi-topological gravities}},\ }\href
  {https://doi.org/10.1007/JHEP11(2019)062} {\bibfield  {journal} {\bibinfo
  {journal} {JHEP}\ }\textbf {\bibinfo {volume} {11}},\ \bibinfo {pages}
  {062}},\ \Eprint {https://arxiv.org/abs/1906.00987} {arXiv:1906.00987
  [hep-th]} \BibitemShut {NoStop}%
\bibitem [{\citenamefont {Bueno}\ \emph {et~al.}(2020)\citenamefont {Bueno},
  \citenamefont {Cano},\ and\ \citenamefont {Hennigar}}]{Bueno:2019ycr}%
  \BibitemOpen
  \bibfield  {author} {\bibinfo {author} {\bibfnamefont {P.}~\bibnamefont
  {Bueno}}, \bibinfo {author} {\bibfnamefont {P.~A.}\ \bibnamefont {Cano}},\
  and\ \bibinfo {author} {\bibfnamefont {R.~A.}\ \bibnamefont {Hennigar}},\
  }\bibfield  {title} {\bibinfo {title} {{(Generalized) quasi-topological
  gravities at all orders}},\ }\href {https://doi.org/10.1088/1361-6382/ab5410}
  {\bibfield  {journal} {\bibinfo  {journal} {Class. Quant. Grav.}\ }\textbf
  {\bibinfo {volume} {37}},\ \bibinfo {pages} {015002} (\bibinfo {year}
  {2020})},\ \Eprint {https://arxiv.org/abs/1909.07983} {arXiv:1909.07983
  [hep-th]} \BibitemShut {NoStop}%
\bibitem [{\citenamefont {Bueno}\ \emph {et~al.}(2023)\citenamefont {Bueno},
  \citenamefont {Cano}, \citenamefont {Hennigar}, \citenamefont {Lu},\ and\
  \citenamefont {Moreno}}]{Bueno:2022res}%
  \BibitemOpen
  \bibfield  {author} {\bibinfo {author} {\bibfnamefont {P.}~\bibnamefont
  {Bueno}}, \bibinfo {author} {\bibfnamefont {P.~A.}\ \bibnamefont {Cano}},
  \bibinfo {author} {\bibfnamefont {R.~A.}\ \bibnamefont {Hennigar}}, \bibinfo
  {author} {\bibfnamefont {M.}~\bibnamefont {Lu}},\ and\ \bibinfo {author}
  {\bibfnamefont {J.}~\bibnamefont {Moreno}},\ }\bibfield  {title} {\bibinfo
  {title} {{Generalized quasi-topological gravities: the whole shebang}},\
  }\href {https://doi.org/10.1088/1361-6382/aca236} {\bibfield  {journal}
  {\bibinfo  {journal} {Class. Quant. Grav.}\ }\textbf {\bibinfo {volume}
  {40}},\ \bibinfo {pages} {015004} (\bibinfo {year} {2023})},\ \Eprint
  {https://arxiv.org/abs/2203.05589} {arXiv:2203.05589 [hep-th]} \BibitemShut
  {NoStop}%
\bibitem [{\citenamefont {Bueno}\ \emph
  {et~al.}(2025{\natexlab{b}})\citenamefont {Bueno}, \citenamefont {Cano},
  \citenamefont {Hennigar},\ and\ \citenamefont {Murcia}}]{Bueno:2024zsx}%
  \BibitemOpen
  \bibfield  {author} {\bibinfo {author} {\bibfnamefont {P.}~\bibnamefont
  {Bueno}}, \bibinfo {author} {\bibfnamefont {P.~A.}\ \bibnamefont {Cano}},
  \bibinfo {author} {\bibfnamefont {R.~A.}\ \bibnamefont {Hennigar}},\ and\
  \bibinfo {author} {\bibfnamefont {{\'A}.~J.}\ \bibnamefont {Murcia}},\
  }\bibfield  {title} {\bibinfo {title} {{Regular black holes from thin-shell
  collapse}},\ }\href {https://doi.org/10.1103/PhysRevD.111.104009} {\bibfield
  {journal} {\bibinfo  {journal} {Phys. Rev. D}\ }\textbf {\bibinfo {volume}
  {111}},\ \bibinfo {pages} {104009} (\bibinfo {year} {2025}{\natexlab{b}})},\
  \Eprint {https://arxiv.org/abs/2412.02740} {arXiv:2412.02740 [gr-qc]}
  \BibitemShut {NoStop}%
\bibitem [{\citenamefont {Konoplya}\ and\ \citenamefont
  {Zhidenko}(2024)}]{Konoplya:2024kih}%
  \BibitemOpen
  \bibfield  {author} {\bibinfo {author} {\bibfnamefont {R.~A.}\ \bibnamefont
  {Konoplya}}\ and\ \bibinfo {author} {\bibfnamefont {A.}~\bibnamefont
  {Zhidenko}},\ }\bibfield  {title} {\bibinfo {title} {{Dymnikova black hole
  from an infinite tower of higher-curvature corrections}},\ }\href
  {https://doi.org/10.1016/j.physletb.2024.138945} {\bibfield  {journal}
  {\bibinfo  {journal} {Phys. Lett. B}\ }\textbf {\bibinfo {volume} {856}},\
  \bibinfo {pages} {138945} (\bibinfo {year} {2024})},\ \Eprint
  {https://arxiv.org/abs/2404.09063} {arXiv:2404.09063 [gr-qc]} \BibitemShut
  {NoStop}%
\bibitem [{\citenamefont {Bueno}\ \emph
  {et~al.}(2026{\natexlab{a}})\citenamefont {Bueno}, \citenamefont {Hennigar},\
  and\ \citenamefont {Murcia}}]{Bueno:2025qjk}%
  \BibitemOpen
  \bibfield  {author} {\bibinfo {author} {\bibfnamefont {P.}~\bibnamefont
  {Bueno}}, \bibinfo {author} {\bibfnamefont {R.~A.}\ \bibnamefont
  {Hennigar}},\ and\ \bibinfo {author} {\bibfnamefont {{\'A}.~J.}\ \bibnamefont
  {Murcia}},\ }\bibfield  {title} {\bibinfo {title} {{Birkhoff implies
  quasi-topological}},\ }\href {https://doi.org/10.1088/1361-6382/ae6600}
  {\bibfield  {journal} {\bibinfo  {journal} {Class. Quant. Grav.}\ }\textbf
  {\bibinfo {volume} {43}},\ \bibinfo {pages} {095020} (\bibinfo {year}
  {2026}{\natexlab{a}})},\ \Eprint {https://arxiv.org/abs/2510.25823}
  {arXiv:2510.25823 [gr-qc]} \BibitemShut {NoStop}%
\bibitem [{\citenamefont {Bueno}\ \emph
  {et~al.}(2026{\natexlab{b}})\citenamefont {Bueno}, \citenamefont {Hennigar},
  \citenamefont {Murcia},\ and\ \citenamefont {Vicente-Cano}}]{Bueno:2025tli}%
  \BibitemOpen
  \bibfield  {author} {\bibinfo {author} {\bibfnamefont {P.}~\bibnamefont
  {Bueno}}, \bibinfo {author} {\bibfnamefont {R.~A.}\ \bibnamefont {Hennigar}},
  \bibinfo {author} {\bibfnamefont {{\'A}.~J.}\ \bibnamefont {Murcia}},\ and\
  \bibinfo {author} {\bibfnamefont {A.}~\bibnamefont {Vicente-Cano}},\
  }\bibfield  {title} {\bibinfo {title} {{Buchdahl limits in theories with
  regular black holes}},\ }\href {https://doi.org/10.1103/mcrq-6fl3} {\bibfield
   {journal} {\bibinfo  {journal} {Phys. Rev. D}\ }\textbf {\bibinfo {volume}
  {113}},\ \bibinfo {pages} {084008} (\bibinfo {year} {2026}{\natexlab{b}})},\
  \Eprint {https://arxiv.org/abs/2512.19796} {arXiv:2512.19796 [gr-qc]}
  \BibitemShut {NoStop}%
\bibitem [{\citenamefont {Ling}\ and\ \citenamefont {Yu}(2026)}]{Ling:2025ncw}%
  \BibitemOpen
  \bibfield  {author} {\bibinfo {author} {\bibfnamefont {Y.}~\bibnamefont
  {Ling}}\ and\ \bibinfo {author} {\bibfnamefont {Z.}~\bibnamefont {Yu}},\
  }\bibfield  {title} {\bibinfo {title} {{Big bounce and black bounce in
  quasi-topological gravity}},\ }\href
  {https://doi.org/10.1088/1475-7516/2026/03/004} {\bibfield  {journal}
  {\bibinfo  {journal} {JCAP}\ }\textbf {\bibinfo {volume} {03}},\ \bibinfo
  {pages} {004}},\ \Eprint {https://arxiv.org/abs/2509.00137} {arXiv:2509.00137
  [gr-qc]} \BibitemShut {NoStop}%
\bibitem [{\citenamefont {Frolov}(2026)}]{Frolov:2025ddw}%
  \BibitemOpen
  \bibfield  {author} {\bibinfo {author} {\bibfnamefont {V.~P.}\ \bibnamefont
  {Frolov}},\ }\bibfield  {title} {\bibinfo {title} {{Quasitopological gravity
  and double-copy formalism}},\ }\href {https://doi.org/10.1103/mbdj-tqtv}
  {\bibfield  {journal} {\bibinfo  {journal} {Phys. Rev. D}\ }\textbf {\bibinfo
  {volume} {113}},\ \bibinfo {pages} {064023} (\bibinfo {year} {2026})},\
  \Eprint {https://arxiv.org/abs/2512.14674} {arXiv:2512.14674 [gr-qc]}
  \BibitemShut {NoStop}%
\bibitem [{\citenamefont {Frolov}\ and\ \citenamefont
  {Zelnikov}(2026)}]{Frolov:2026rcm}%
  \BibitemOpen
  \bibfield  {author} {\bibinfo {author} {\bibfnamefont {V.~P.}\ \bibnamefont
  {Frolov}}\ and\ \bibinfo {author} {\bibfnamefont {A.}~\bibnamefont
  {Zelnikov}},\ }\bibfield  {title} {\bibinfo {title} {{Regular black holes in
  quasitopological gravity: Null shells and mass inflation}},\ }\href
  {https://doi.org/10.1103/nrd7-j6hs} {\bibfield  {journal} {\bibinfo
  {journal} {Phys. Rev. D}\ }\textbf {\bibinfo {volume} {113}},\ \bibinfo
  {pages} {084007} (\bibinfo {year} {2026})},\ \Eprint
  {https://arxiv.org/abs/2601.01861} {arXiv:2601.01861 [gr-qc]} \BibitemShut
  {NoStop}%
\bibitem [{\citenamefont {Deser}\ \emph {et~al.}(2008)\citenamefont {Deser},
  \citenamefont {Sarioglu},\ and\ \citenamefont {Tekin}}]{Deser:2007za}%
  \BibitemOpen
  \bibfield  {author} {\bibinfo {author} {\bibfnamefont {S.}~\bibnamefont
  {Deser}}, \bibinfo {author} {\bibfnamefont {O.}~\bibnamefont {Sarioglu}},\
  and\ \bibinfo {author} {\bibfnamefont {B.}~\bibnamefont {Tekin}},\ }\bibfield
   {title} {\bibinfo {title} {{Spherically symmetric solutions of Einstein +
  non-polynomial gravities}},\ }\href
  {https://doi.org/10.1007/s10714-007-0508-1} {\bibfield  {journal} {\bibinfo
  {journal} {Gen. Rel. Grav.}\ }\textbf {\bibinfo {volume} {40}},\ \bibinfo
  {pages} {1} (\bibinfo {year} {2008})},\ \Eprint
  {https://arxiv.org/abs/0705.1669} {arXiv:0705.1669 [gr-qc]} \BibitemShut
  {NoStop}%
\bibitem [{\citenamefont {Gao}(2012)}]{Gao:2012fd}%
  \BibitemOpen
  \bibfield  {author} {\bibinfo {author} {\bibfnamefont {C.}~\bibnamefont
  {Gao}},\ }\bibfield  {title} {\bibinfo {title} {{Generalized modified gravity
  with the second order acceleration equation}},\ }\href
  {https://doi.org/10.1103/PhysRevD.86.103512} {\bibfield  {journal} {\bibinfo
  {journal} {Phys. Rev. D}\ }\textbf {\bibinfo {volume} {86}},\ \bibinfo
  {pages} {103512} (\bibinfo {year} {2012})},\ \Eprint
  {https://arxiv.org/abs/1208.2790} {arXiv:1208.2790 [gr-qc]} \BibitemShut
  {NoStop}%
\bibitem [{\citenamefont {Coll{\'e}aux}\ and\ \citenamefont
  {Zerbini}(2015)}]{Colleaux:2015yta}%
  \BibitemOpen
  \bibfield  {author} {\bibinfo {author} {\bibfnamefont {A.}~\bibnamefont
  {Coll{\'e}aux}}\ and\ \bibinfo {author} {\bibfnamefont {s.}~\bibnamefont
  {Zerbini}},\ }\bibfield  {title} {\bibinfo {title} {{Modified Gravity Models
  Admitting Second Order Equations of Motion}},\ }\href
  {https://doi.org/10.3390/e17106643} {\bibfield  {journal} {\bibinfo
  {journal} {Entropy}\ }\textbf {\bibinfo {volume} {17}},\ \bibinfo {pages}
  {6643} (\bibinfo {year} {2015})},\ \Eprint {https://arxiv.org/abs/1508.06178}
  {arXiv:1508.06178 [gr-qc]} \BibitemShut {NoStop}%
\bibitem [{\citenamefont {Coll{\'e}aux}\ \emph {et~al.}(2018)\citenamefont
  {Coll{\'e}aux}, \citenamefont {Chinaglia},\ and\ \citenamefont
  {Zerbini}}]{Colleaux:2017ibe}%
  \BibitemOpen
  \bibfield  {author} {\bibinfo {author} {\bibfnamefont {A.}~\bibnamefont
  {Coll{\'e}aux}}, \bibinfo {author} {\bibfnamefont {S.}~\bibnamefont
  {Chinaglia}},\ and\ \bibinfo {author} {\bibfnamefont {S.}~\bibnamefont
  {Zerbini}},\ }\bibfield  {title} {\bibinfo {title} {{Nonpolynomial Lagrangian
  approach to regular black holes}},\ }\href
  {https://doi.org/10.1142/S0218271818300021} {\bibfield  {journal} {\bibinfo
  {journal} {Int. J. Mod. Phys. D}\ }\textbf {\bibinfo {volume} {27}},\
  \bibinfo {pages} {1830002} (\bibinfo {year} {2018})},\ \Eprint
  {https://arxiv.org/abs/1712.03730} {arXiv:1712.03730 [gr-qc]} \BibitemShut
  {NoStop}%
\bibitem [{\citenamefont {Colleaux}(2019)}]{Colleaux:2019ckh}%
  \BibitemOpen
  \bibfield  {author} {\bibinfo {author} {\bibfnamefont {A.}~\bibnamefont
  {Colleaux}},\ }\emph {\bibinfo {title} {{Regular black hole and cosmological
  spacetimes in Non-Polynomial Gravity theories}}},\ \href@noop {} {Ph.D.
  thesis},\ \bibinfo  {school} {Trento U.} (\bibinfo {year} {2019})\BibitemShut
  {NoStop}%
\bibitem [{\citenamefont {Bueno}\ \emph
  {et~al.}(2026{\natexlab{c}})\citenamefont {Bueno}, \citenamefont {Cano},
  \citenamefont {Hennigar},\ and\ \citenamefont {Murcia}}]{Bueno:2025zaj}%
  \BibitemOpen
  \bibfield  {author} {\bibinfo {author} {\bibfnamefont {P.}~\bibnamefont
  {Bueno}}, \bibinfo {author} {\bibfnamefont {P.~A.}\ \bibnamefont {Cano}},
  \bibinfo {author} {\bibfnamefont {R.~A.}\ \bibnamefont {Hennigar}},\ and\
  \bibinfo {author} {\bibfnamefont {{\'A}.~J.}\ \bibnamefont {Murcia}},\
  }\bibfield  {title} {\bibinfo {title} {{Regular black hole formation in
  four-dimensional nonpolynomial gravities}},\ }\href
  {https://doi.org/10.1103/8f3j-zcxh} {\bibfield  {journal} {\bibinfo
  {journal} {Phys. Rev. D}\ }\textbf {\bibinfo {volume} {113}},\ \bibinfo
  {pages} {024019} (\bibinfo {year} {2026}{\natexlab{c}})},\ \Eprint
  {https://arxiv.org/abs/2509.19016} {arXiv:2509.19016 [gr-qc]} \BibitemShut
  {NoStop}%
\bibitem [{\citenamefont {Tan}\ and\ \citenamefont {Wang}(2026)}]{Tan:2025hht}%
  \BibitemOpen
  \bibfield  {author} {\bibinfo {author} {\bibfnamefont {C.}~\bibnamefont
  {Tan}}\ and\ \bibinfo {author} {\bibfnamefont {Y.-Q.}\ \bibnamefont {Wang}},\
  }\bibfield  {title} {\bibinfo {title} {{Frozen neutron stars in
  four-dimensional nonpolynomial gravities}},\ }\href
  {https://doi.org/10.1103/jmnh-p43b} {\bibfield  {journal} {\bibinfo
  {journal} {Phys. Rev. D}\ }\textbf {\bibinfo {volume} {114}},\ \bibinfo
  {pages} {064007} (\bibinfo {year} {2026})},\ \Eprint
  {https://arxiv.org/abs/2512.23525} {arXiv:2512.23525 [gr-qc]} \BibitemShut
  {NoStop}%
\bibitem [{\citenamefont {Borissova}\ and\ \citenamefont
  {Carballo-Rubio}(2026)}]{Borissova:2026wmn}%
  \BibitemOpen
  \bibfield  {author} {\bibinfo {author} {\bibfnamefont {J.}~\bibnamefont
  {Borissova}}\ and\ \bibinfo {author} {\bibfnamefont {R.}~\bibnamefont
  {Carballo-Rubio}},\ }\bibfield  {title} {\bibinfo {title} {{Regular black
  holes from pure gravity in four dimensions}},\ }\href
  {https://doi.org/10.1103/6x2z-qbkh} {\bibfield  {journal} {\bibinfo
  {journal} {Phys. Rev. D}\ }\textbf {\bibinfo {volume} {113}},\ \bibinfo
  {pages} {124004} (\bibinfo {year} {2026})},\ \Eprint
  {https://arxiv.org/abs/2602.16773} {arXiv:2602.16773 [gr-qc]} \BibitemShut
  {NoStop}%
\bibitem [{\citenamefont {Einstein}(1936)}]{Einstein:1936llh}%
  \BibitemOpen
  \bibfield  {author} {\bibinfo {author} {\bibfnamefont {A.}~\bibnamefont
  {Einstein}},\ }\bibfield  {title} {\bibinfo {title} {{Lens-Like Action of a
  Star by the Deviation of Light in the Gravitational Field}},\ }\href
  {https://doi.org/10.1126/science.84.2188.506} {\bibfield  {journal} {\bibinfo
   {journal} {Science}\ }\textbf {\bibinfo {volume} {84}},\ \bibinfo {pages}
  {506} (\bibinfo {year} {1936})}\BibitemShut {NoStop}%
\bibitem [{\citenamefont {Chen}\ and\ \citenamefont
  {Jing}(2015)}]{Chen:2015cpa}%
  \BibitemOpen
  \bibfield  {author} {\bibinfo {author} {\bibfnamefont {S.}~\bibnamefont
  {Chen}}\ and\ \bibinfo {author} {\bibfnamefont {J.}~\bibnamefont {Jing}},\
  }\bibfield  {title} {\bibinfo {title} {{Strong gravitational lensing for the
  photons coupled to Weyl tensor in a Schwarzschild black hole spacetime}},\
  }\href {https://doi.org/10.1088/1475-7516/2015/10/002} {\bibfield  {journal}
  {\bibinfo  {journal} {JCAP}\ }\textbf {\bibinfo {volume} {10}},\ \bibinfo
  {pages} {002}},\ \Eprint {https://arxiv.org/abs/1502.01088} {arXiv:1502.01088
  [gr-qc]} \BibitemShut {NoStop}%
\bibitem [{\citenamefont {Lu}\ \emph {et~al.}(2016)\citenamefont {Lu},
  \citenamefont {Yang},\ and\ \citenamefont {Xie}}]{Lu:2016gsf}%
  \BibitemOpen
  \bibfield  {author} {\bibinfo {author} {\bibfnamefont {X.}~\bibnamefont
  {Lu}}, \bibinfo {author} {\bibfnamefont {F.-W.}\ \bibnamefont {Yang}},\ and\
  \bibinfo {author} {\bibfnamefont {Y.}~\bibnamefont {Xie}},\ }\bibfield
  {title} {\bibinfo {title} {{Strong gravitational field time delay for photons
  coupled to Weyl tensor in a Schwarzschild black hole}},\ }\href
  {https://doi.org/10.1140/epjc/s10052-016-4218-2} {\bibfield  {journal}
  {\bibinfo  {journal} {Eur. Phys. J. C}\ }\textbf {\bibinfo {volume} {76}},\
  \bibinfo {pages} {357} (\bibinfo {year} {2016})},\ \Eprint
  {https://arxiv.org/abs/1606.02932} {arXiv:1606.02932 [gr-qc]} \BibitemShut
  {NoStop}%
\bibitem [{\citenamefont {Chen}\ \emph {et~al.}(2017)\citenamefont {Chen},
  \citenamefont {Wang}, \citenamefont {Huang}, \citenamefont {Jing},\ and\
  \citenamefont {Wang}}]{Chen:2016hil}%
  \BibitemOpen
  \bibfield  {author} {\bibinfo {author} {\bibfnamefont {S.}~\bibnamefont
  {Chen}}, \bibinfo {author} {\bibfnamefont {S.}~\bibnamefont {Wang}}, \bibinfo
  {author} {\bibfnamefont {Y.}~\bibnamefont {Huang}}, \bibinfo {author}
  {\bibfnamefont {J.}~\bibnamefont {Jing}},\ and\ \bibinfo {author}
  {\bibfnamefont {S.}~\bibnamefont {Wang}},\ }\bibfield  {title} {\bibinfo
  {title} {{Strong gravitational lensing for the photons coupled to a Weyl
  tensor in a Kerr black hole spacetime}},\ }\href
  {https://doi.org/10.1103/PhysRevD.95.104017} {\bibfield  {journal} {\bibinfo
  {journal} {Phys. Rev. D}\ }\textbf {\bibinfo {volume} {95}},\ \bibinfo
  {pages} {104017} (\bibinfo {year} {2017})},\ \Eprint
  {https://arxiv.org/abs/1611.08783} {arXiv:1611.08783 [gr-qc]} \BibitemShut
  {NoStop}%
\bibitem [{\citenamefont {Dymnikova}(1992)}]{Dymnikova:1992ux}%
  \BibitemOpen
  \bibfield  {author} {\bibinfo {author} {\bibfnamefont {I.}~\bibnamefont
  {Dymnikova}},\ }\bibfield  {title} {\bibinfo {title} {{Vacuum nonsingular
  black hole}},\ }\href {https://doi.org/10.1007/BF00760226} {\bibfield
  {journal} {\bibinfo  {journal} {Gen. Rel. Grav.}\ }\textbf {\bibinfo {volume}
  {24}},\ \bibinfo {pages} {235} (\bibinfo {year} {1992})}\BibitemShut
  {NoStop}%
\bibitem [{\citenamefont {Kanai}(2026{\natexlab{b}})}]{Kanai:2026bag}%
  \BibitemOpen
  \bibfield  {author} {\bibinfo {author} {\bibfnamefont {T.}~\bibnamefont
  {Kanai}},\ }\bibfield  {title} {\bibinfo {title} {{Gravitational Lensing of
  Hayward Black Holes with EFT-Corrected Photon Propagation}},\ }\href@noop {}
  {\  (\bibinfo {year} {2026}{\natexlab{b}})},\ \Eprint
  {https://arxiv.org/abs/2608.26785} {arXiv:2608.26785 [gr-qc]} \BibitemShut
  {NoStop}%
\bibitem [{\citenamefont {Kanai}(2026{\natexlab{c}})}]{Kanai:2026xpw}%
  \BibitemOpen
  \bibfield  {author} {\bibinfo {author} {\bibfnamefont {T.}~\bibnamefont
  {Kanai}},\ }\bibfield  {title} {\bibinfo {title} {{Photon Surfaces in
  Higher-Curvature Gravity: Implications for Quasinormal Modes and
  Gravitational Lensing}},\ }\href@noop {} {\  (\bibinfo {year}
  {2026}{\natexlab{c}})},\ \Eprint {https://arxiv.org/abs/2604.24115}
  {arXiv:2604.24115 [gr-qc]} \BibitemShut {NoStop}%
\end{thebibliography}%

\end{document}